\documentclass[iop, apj, twocolappendix, numberedappendix, appendixfloats]{emulateapj}
\usepackage{threeparttable}
\usepackage{multirow}
\usepackage{times}
\usepackage{enumitem}
\usepackage{url} 
\usepackage{amsmath,bm}

\shorttitle{}
\shortauthors{Y. Huang et al.}

\begin{document}

\title{Beyond spectroscopy. I. Metallicities, distances, and age estimates for over twenty million stars from  SMSS DR2 and Gaia EDR3}
\author{Yang Huang\altaffilmark{1}}
\author{Timothy C. Beers\altaffilmark{2}}
\author{Christian Wolf\altaffilmark{3,4}}
\author{Young Sun Lee\altaffilmark{5}}
\author{Christopher A. Onken\altaffilmark{3,4}}
\author{Haibo Yuan\altaffilmark{6}}
\author{Derek Shank\altaffilmark{2}}
\author{Huawei Zhang\altaffilmark{7,8}}
\author{Chun Wang\altaffilmark{9}}
\author{Jianrong Shi\altaffilmark{10}}
\author{Zhou Fan\altaffilmark{10}}

\altaffiltext{1}{South-Western Institute for Astronomy Research, Yunnan University, Kunming 650500, People's Republic of China; {\it yanghuang@ynu.edu.cn}}
\altaffiltext{2}{Department of Physics and JINA Center for the Evolution of the Elements (JINA-CEE), University of Notre Dame, Notre Dame, IN 46556, USA}
\altaffiltext{3}{Research School of Astronomy and Astrophysics, Australian National University, Canberra, ACT 2611, Australia}
\altaffiltext{4}{Centre for Gravitational Astrophysics, Research Schools of Physics, and Astronomy and Astrophysics, Australian National University, Canberra, ACT 2611, Australia}
\altaffiltext{5}{Department of Astronomy and Space Science, Chungnam National University, Daejeon 34134, Republic of Korea}
\altaffiltext{6}{Department of Astronomy, Beijing Normal University, Beijing 100875, People's Republic of China}
\altaffiltext{7}{Department of Astronomy, School of Physics, Peking University, Beijing 100871, People's Republic of China}
\altaffiltext{8}{Kavli Institute for Astronomy and Astrophysics, Peking University, Beijing 100871, People's Republic of China}
\altaffiltext{9}{Tianjin Astrophysics Center, Tianjin Normal University, Tianjin 300387, People's Republic of China}
\altaffiltext{10}{Key Lab of Optical Astronomy, National Astronomical Observatories, Chinese Academy of Sciences, Beijing 100012, People's Republic of China}

\begin{abstract}
Accurate determinations of stellar parameters and distances for large complete samples of stars are keys for conducting detailed studies of the formation and evolution of our Galaxy. Here we present stellar atmospheric parameters (effective temperature, luminosity classifications, and metallicity) estimates for some 24\,million stars determined from the stellar colors of SMSS\,DR2 and {\it\,Gaia}\,EDR3, based on training datasets with available spectroscopic measurements from previous high/medium/low-resolution spectroscopic surveys. The number of stars with photometric-metallicity estimates is 4--5\,times larger than that collected by the current largest spectroscopic survey to date\,--\,LAMOST\,--\,over the course of the past decade. External checks indicate that the precision of the photometric-metallicity estimates are quite high, comparable to or slightly better than that derived from spectroscopy, with typical values around 0.05--0.15\,dex for both dwarf and giant stars with [Fe/H]\,$>-1.0$, 0.10--0.20\,dex for giant stars with $-2.0<$\,[Fe/H]\,$\le-1.0$. and 0.20--0.25\,dex for giant stars with [Fe/H]\,$\le-2.0$, and include estimates for stars as metal-poor as [Fe/H] $\sim-3.5$, substantially lower than previous photometric techniques. Photometric-metallicity estimates are obtained for an unprecedented number of metal-poor stars, including  a total of over three million metal-poor (MP;\,[Fe/H]\,$\le-1.0$) stars, over half a million very metal-poor (VMP;\,[Fe/H]\,$\le-2.0$) stars, and over 25,000\,extremely metal-poor (EMP;\,[Fe/H]\,$\le\,-3.0$) stars. Moreover, distances are determined for over 20\,million stars in our sample. For the over 18\,million sample stars with accurate {\it\,Gaia} parallaxes, stellar ages are estimated by comparing with theoretical isochrones.  Astrometric information is provided for the stars in our catalog, along with radial velocities for $\sim10$\% of our sample stars, taken from completed/ongoing large-scale spectroscopic surveys.
\end{abstract}

\section{Introduction}

\begin{table*}
\centering
\caption{Ongoing and Planned Narrow/Medium-bandwidth Large-Scale Photometric Surveys}
\begin{threeparttable}
\begin{tabular}{ccccccc}
\hline
Survey&Aperture&Field of view&Sky area (N/S)\tnote{a}&$N_{\rm N/M}$\tnote{b}&Depth ($5\sigma$)& Status\\
&(meters)&(square degrees)&(square degrees)&&&\\ 
\hline
\hline
SkyMapper&1.35&5.7& 21,360 S&2&$\sim 21.7$ at $r$-band&Ongoing\\
Pristine&3.6&1.0&$> $2500 N&1&$\sim 21.0$ at $g$-band ($10\sigma$)&Ongoing\\
SAGE&2.3/1/1\tnote{c}&1.1/2.3/0.4\tnote{c}&12,000 N&5&$\sim 20$ at $V$-band&Ongoing\\
J-PLUS&0.8&2.0&8500 N&7&$\sim 21.5$ at $r$-band&Ongoing\\
S-PLUS&0.8&2.0&9300 S&7&$\sim 21.3$ at $r$-band ($3\sigma$)&Ongoing\\
J-PAS&2.5&4.7&8500 N&54&$\sim 24.0$ at $r$-band&Ongoing\\
Mephisto&1.6&2.1&$>$20,000 N&2&$\sim 23.3$ at $r$-band&Planned\\
\hline 
\end{tabular}
\begin{tablenotes}
\item[a]Here N  and S represent the Northern and Southern Hemispheres, respectively.
\item[b]The number of narrow/medium-bandwidth filters adopted by this survey.
\item[c]The SAGE survey is based on three telescopes: the Bok 2.3m Telescope for $uv$ filters, the Nanshan 1m Wide-field Telescope for $gri$ filters, and the Zeiss 1m Telescope at the Maidanak Astronomical  Observatory for H$\alpha$ narrow and wide-band filters.
\end{tablenotes}
\end{threeparttable}
\end{table*}

The field of Galactic Archaeology has entered a golden era, due to the culmination of decades of large-scale spectroscopic efforts such as the HK Survey (Beers, Preston, \& Shectman 1985, 1992), the Hamburg/ESO Survey (HES; Christlieb 2003), the Sloan Digital Sky Survey (SDSS; York et al. 2000), the Radial Velocity Experiment (RAVE; Steinmetz et al. 2006), the Sloan Extension for Galactic Understanding and Exploration (SEGUE; Yanny et al. 2009), the Large Sky Area Multi-Object Fiber Spectroscopic Telescope (LAMOST; Deng et al. 2012; Liu et al. 2014), the Galactic Archaeology with HERMES project (GALAH; De Silva et al. 2015), the Apache Point Observatory Galactic Evolution Experiment (APOGEE; Majewski et al. 2017), and the
Hectochelle in the Halo at High-Resolution survey (H3; Conroy et al. 2019).  Stars covering an enormous range of metallicity, including large numbers of stars with metallicities below the lowest-abundance globular clusters (stars once thought not to exist, on theoretical grounds, as recently as the early 1980s), have been discovered and analyzed in great detail (see reviews by Beers \& Christlieb 2005; Ivez{\'i}c, Beers, \& Juri{\'c} 2012; Frebel \& Norris 2015). Putting these discoveries into the context of the stellar populations of the Galaxy has been expedited greatly by the successful {\it Gaia} mission and its data releases to date (Gaia Collaboration 2016, 2018, 2021). These surveys, collectively, have enabled astronomers to draw a much clearer picture of the stellar populations of our Milky Way (MW), and significantly advanced our knowledge of its chemical evolution and assembly history.

 The location of the Sun in the disk of the MW presents a challenge for the task of obtaining an unbiased, representative sample of stars with available full multi-dimensional information (stellar abundances, distances, motions, and ages), for a number of reasons.  For one, the selection functions of spectroscopic surveys to date are all different from one another, and often complex. Even {\it Gaia} Early Data Release 3 (hereafter EDR3; Gaia Collaboration et al. 2021), which has sampled one percent of the stars of the MW (several billion stars), still exhibits significant spatial patterns due to its observational strategy.  The sampling of Galactic spectroscopic surveys is even more sparse.  For example, the LAMOST Galactic spectroscopic survey (the largest to date, which has been underway for almost a decade), has collected over 8 million spectra with signal-to-noise ratio (SNR) greater than 10 for over 4 million unique stars in its latest public data release\footnote{http://dr6.lamost.org/}. Although ongoing and next-generation spectroscopic surveys will greatly expand the numbers of stars examined in the MW, they will still pale in raw numbers of stars compared to the large-scale astrometric surveys such as $Gaia$, and of course, will still have to deal with the impact of their target-selection criteria in order to extract knowledge and understanding from their data.

On the other hand, the large-scale ongoing and planned narrow/medium-bandwidth photometric surveys (see Table\,1 for details), such as the SkyMapper Southern Survey (SMSS; Wolf et al. 2018), the Pristine survey (Starkenburg et al. 2017), the Stellar Abundance and Galactic
Evolution survey (SAGE; Zheng et al. 2018), the Javalambre Physics of the Accelerating Universe Astrophysical Survey (J-PAS; Benitez et al. 2014), the Javalambre/Southern Photometric Local Universe Survey (J/S-PLUS; Cenarro et al. 2019; Mendes de Oliveira 2019), and the Multi-channel Photometric Survey Telescope (Mephisto; Er et al., in prep.), provide a new way to alleviate the target-selection bias and sparse sampling issues of the current large-scale Galactic spectroscopic surveys.
The narrow/medium-bands adopted by those surveys provide the possibility to perform precise estimates  of stellar atmospheric parameters (e.g., effective temperature, $T_{\rm eff}$, surface gravity, log\,$g$, and metallicity, [Fe/H]), and even some elemental abundances (e.g., Starkenburg et al. 2017; Casagrande et al. 2019; Huang et al. 2019; Whitten et al. 2019, 2021; Yang et al. 2021).
In this way, photometric metallicity and elemental abundances can be estimated for tens of millions to (eventually) billions of stars, approximately matching the size of the astrometric sample achieved by the {\it Gaia} mission.

The advantage of photometric exploration of the MW's stellar populations has been demonstrated by a number of studies in the literature. Juri{\'c} et al. (2008) presented the first panoramic view of the Milky Way, and delineated a three-dimensional number-density distribution of stars from the early data obtained by SDSS imaging data. Following this work, Ivezi{\'c} et al. (2008) developed a photometric-metallicity technique, tagging millions of stars by metallicity, which directly proved the extent of major stellar populations constituting the thin disk, the thick disk, and the halo.  It is worthwhile to point out that this seminal effort suffered from the difficulty that, due to the lack of sensitivity of the broad-band metallicity estimator they employed, their work was unable to consider the presence of stellar populations (in particular, the outer-halo population) with metallicities [Fe/H] $< -2$, later shown to be of importance from spectroscopic samples. These studies have essentially opened up a new era of stellar-population studies, standing on the shoulders of massive imaging-survey data, and have removed the constraints imposed by narrow-angle pencil-beam observations. Recently, An \& Beers (2020, 2021a, 2021b) have taken the photometric mapping of the MW a step further, by combining spatial and chemical information with {\it Gaia} proper motions. Their chemodynamical ``blueprint" shows the full set of recognized substructures in the local volume -- each of these pieces have been studied in depth from previous spectroscopic studies, but they have never been put together to convincingly produce the big picture of the contents and extent of each population. Inspection of their maps clearly illustrates why progress in understanding the nature of stellar populations by ``stiching together" multiple spectroscopic surveys proved so difficult; depending on the cuts adopted by individual researchers in metallicity and distance, very different (and sometimes contradictory) interpretations were unavoidable.

As a pioneering experiment, we here present reliable photometric-metallicity estimates (and other parameters) for over 24 million FGK ($4000 \le T_{\rm eff} \le 6800$\,K) stars from the the second data release  (DR2) of SMSS (Onken et al. 2019; hereafter O19), based on a recalibration of the photometric zero-points provided by Huang et al. (2021).  The number of stars with metallicity estimates is 4--5 times larger than the number obtained by the aforementioned LAMOST spectroscopic survey.  In addition, distance and stellar-age estimates are determined for over 85\% (i.e., over 20 million) and 74\% (i.e., over 18 million) of these stars, respectively. 

The paper is organized as follows. 
In Section\,2, we introduce the data used in the current work.
In Section\,3, determinations of photometric-metallicity estimates based on the recalibrated SMSS DR2 photometry are described in detail, and compared with previous spectroscopic determinations for large numbers of stars in common.  Section\,4 describes estimates of effective temperature, $T_{\rm eff}$.
Distances and stellar-age estimates are presented in Sections\,5 and 6, respectively. 
Radial velocity measurements are collected from previous spectroscopic surveys as described in Section\,7.
We discuss the nature of the full sample, and provide perspectives on future efforts, in Section\,8.  Finally, a summary is presented in Section\,9.\\
\\

\begin{figure}
\begin{center}
\includegraphics[scale=0.325,angle=0]{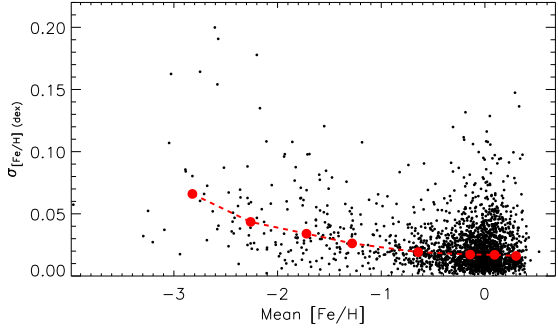}
\caption{Standard deviation of [Fe/H] as a function of [Fe/H] for over 2000 stars with number of metallicity measurements greater than 4 in the PASTEL database.
The red dots represent the median standard deviation of [Fe/H] in the individual mean [Fe/H] bins.
The red dashed line simply connects the dots.}
\end{center}
\end{figure}

\begin{figure*}
\begin{center}
\includegraphics[scale=0.425,angle=0]{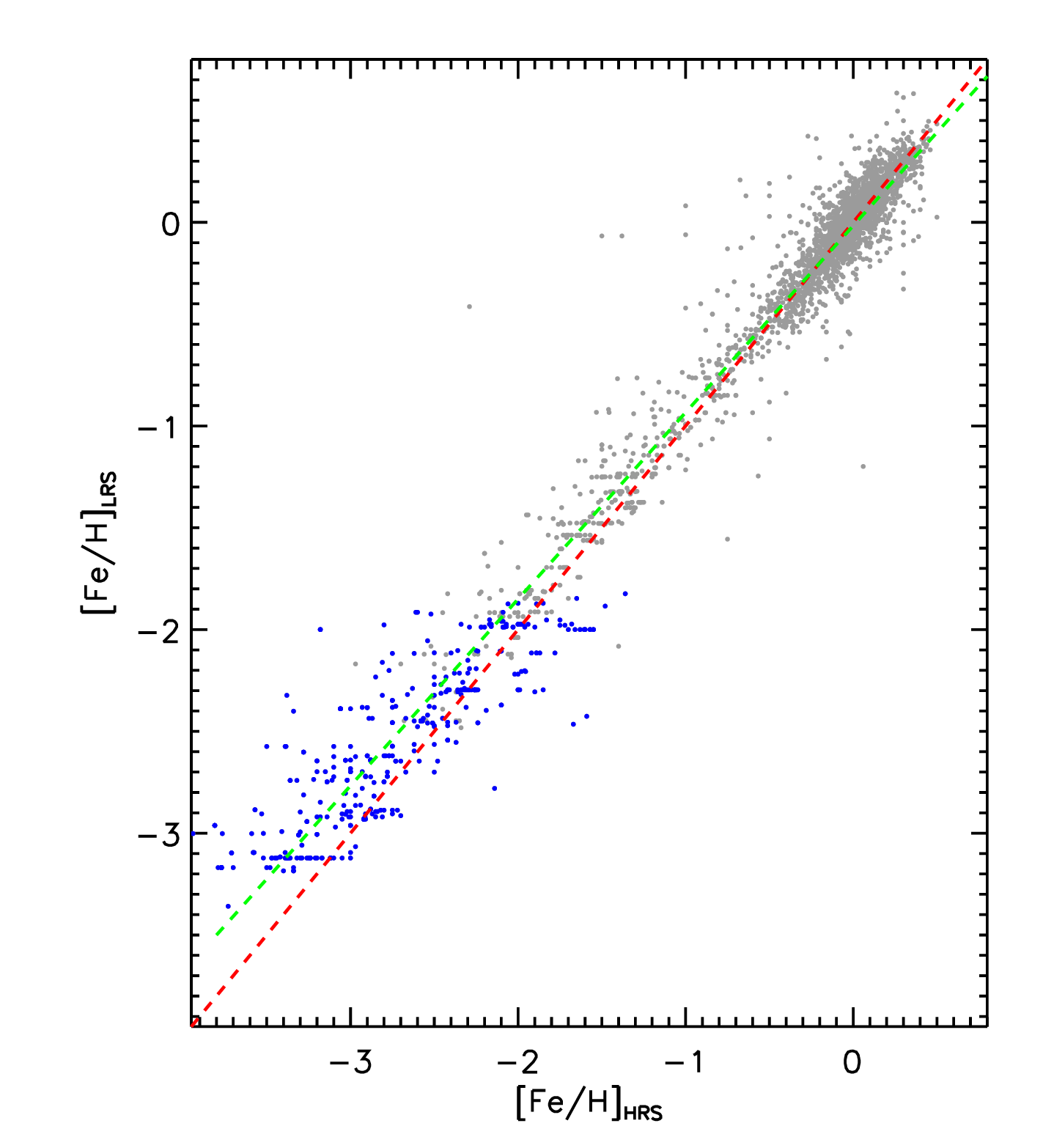}
\includegraphics[scale=0.425,angle=0]{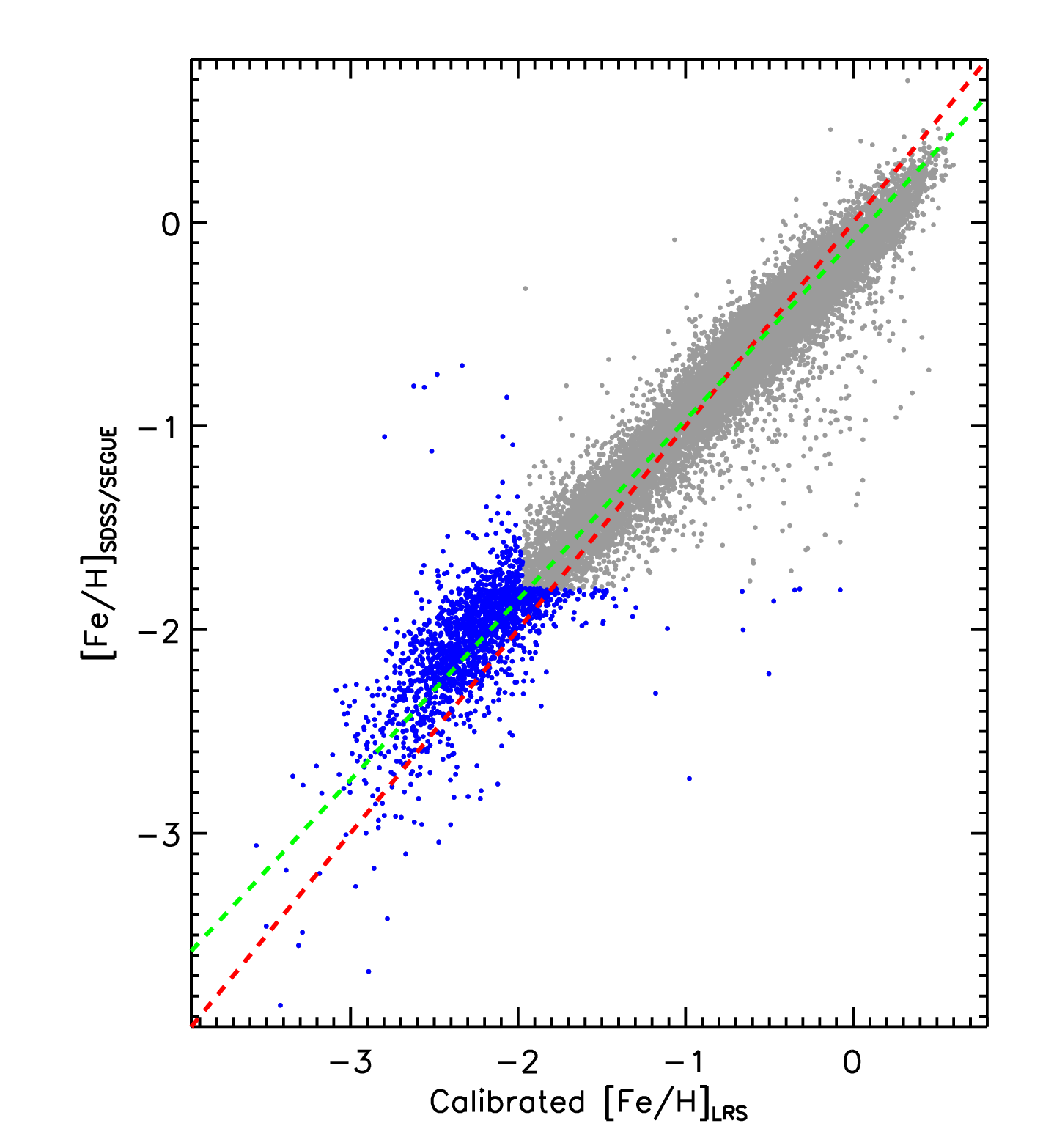}
\caption{{\it Left panel:} Comparison of the metallicity ([Fe/H]) from low-resolution spectra (LRS; $R \sim 1800$)  from LAMOST DR7 with those from the High-Resolution Spectra (HRS, with $R > 20,000$) sample. 
The red dashed line represents the one-to-one line.
The green line represents a first-order polynomial fit of the LRS results to the HRS results. 
Gray and blue dots represent metallicity estimated by the default LAMOST pipeline (LASP) and a custom version of the SSPP (LSSPP; see Lee et al. 2015), optimized for working with LAMOST LRS spectra, respectively.
{\it Right panel:} Comparison of the metallicity ([Fe/H]) for the  SDSS/SEGUE  stars with those from the (calibrated) estimate from LAMOST spectra.
The red dashed line again represents the one-to-one line.
The green line represents a first-order polynomial fit to the SDSS/SEGUE [Fe/H] to the (calibrated) LAMOST LRS estimate.
Gray dots represent metallicity estimated by the LASP and the SSPP, while blue dots represent .metallicity estimated by the LSSPP and recalibrated SSPP.}
\end{center}
\end{figure*}

\begin{figure*}
\begin{center}
\includegraphics[scale=0.35,angle=0]{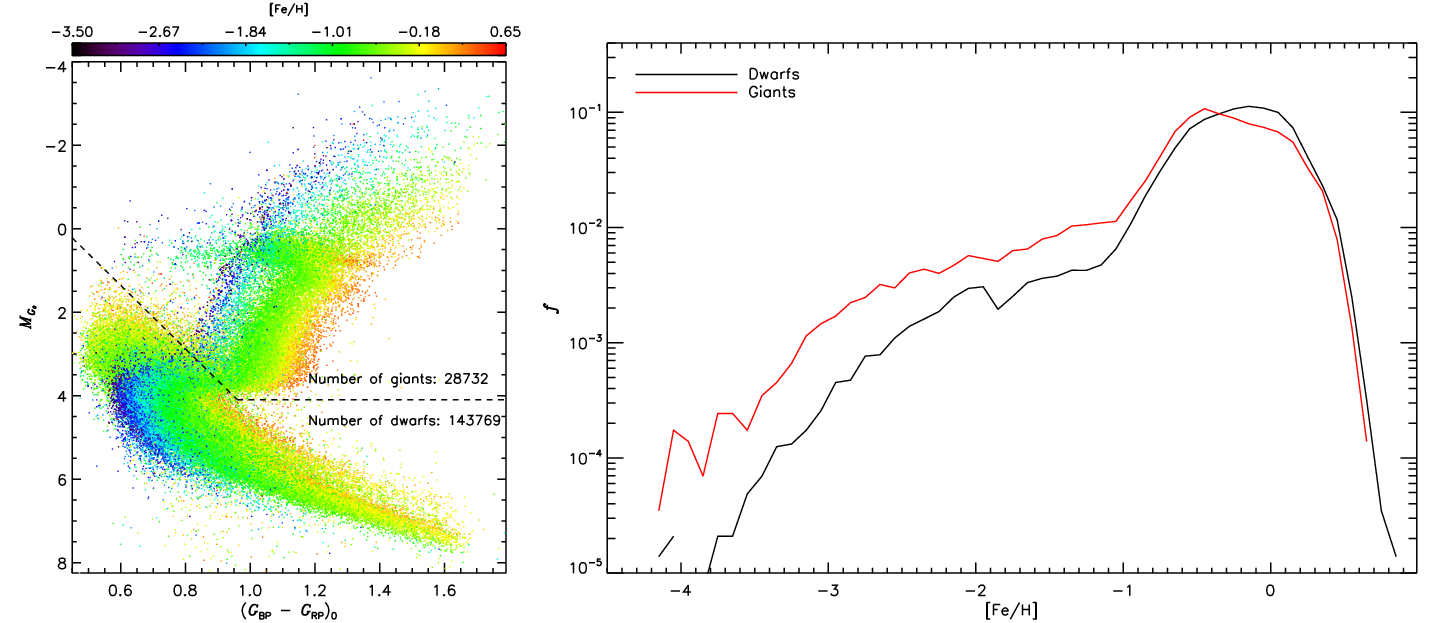}
\caption{{\it Left panel:} Color--absolute magnitude diagram of the training sample defined in Section\,3.1, color-coded by metallicity, as shown in the top color bar. The dashed lines represent the cuts $M_{G_0} = -3.20 + 7.60\cdot (G_{\rm BP} - G_{\rm RP})_0$ or $M_{G_0} = 4.1$, used to separate dwarf and giant stars.
{\it Right panel:} Metallicity ([Fe/H]) distributions of the training sample. Black and red lines mark the dwarf and giant stars, respectively.}
\end{center}
\end{figure*}

\begin{figure*}
\begin{center}
\includegraphics[scale=0.395,angle=0]{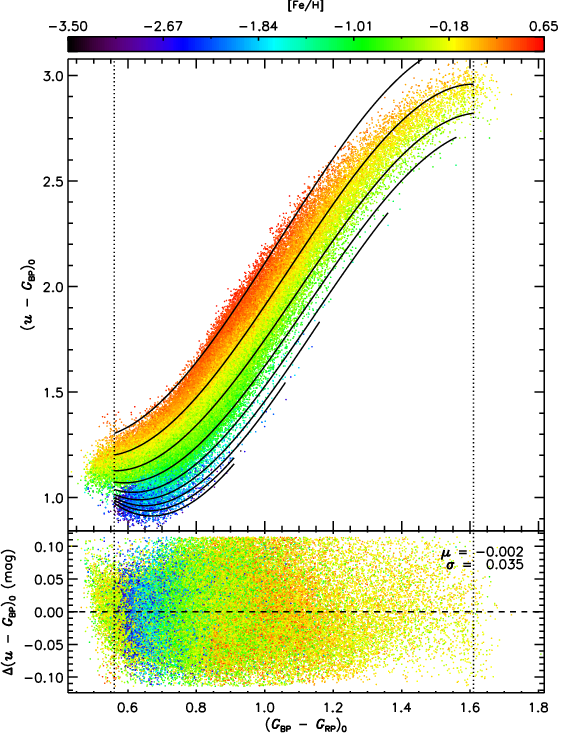}
\includegraphics[scale=0.395,angle=0]{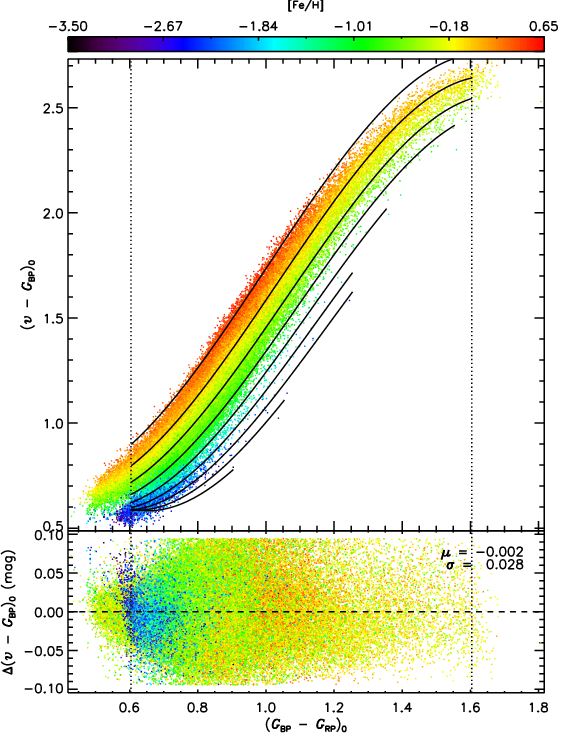}
\caption{Distributions of the training-sample dwarf stars in the $(u - G_{\rm BP})_0$ versus $(G_{\rm BP} - G_{\rm RP})_0$ plane (left panel) and $(v - G_{\rm BP})_0$ versus $(G_{\rm BP} - G_{\rm RP})_0$ plane (right panel), color-coded by metallicity ([Fe/H]), as shown by the top color bars.
The black lines represent our best fits for different values of [Fe/H], as described by Equation\,3.
From top to bottom, the values of [Fe/H] are $+0.5$, 0.0, $-0.5$, $-1.0$, $-1.5$, $-2.0$, $-2.5$, $-3.0$ and $-3.5$, respectively.
The dashed lines mark the color region in $(G_{\rm BP} - G_{\rm RP})_0$ that the fits yield results without systematics.
The lower part of each panel shows the fit residual, as a function of color $(G_{\rm BP} - G_{\rm RP})_0$, with the values of median and standard deviation of the residual marked in the top-right corner.}
\end{center}
\end{figure*}

\begin{figure*}
\begin{center}
\includegraphics[scale=0.395,angle=0]{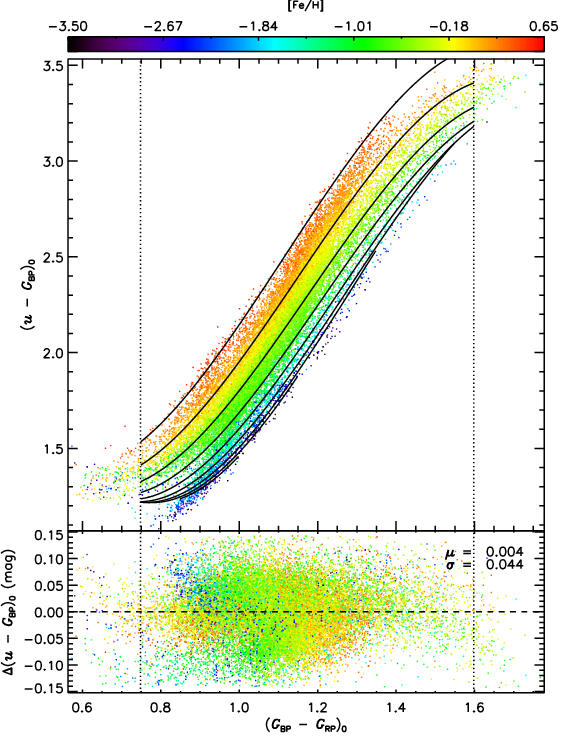}
\includegraphics[scale=0.395,angle=0]{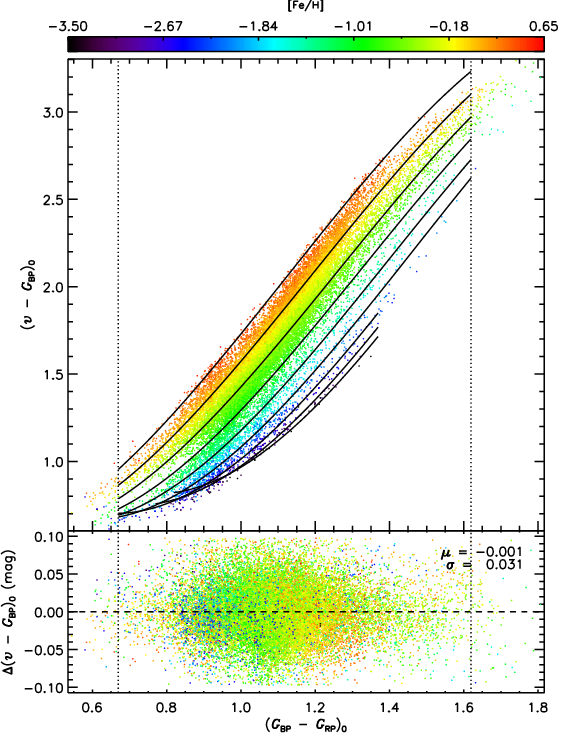}
\caption{Similar to Fig.\,3 but for training-sample giant stars.}
\end{center}
\end{figure*}

\section{Data}
In the current work, we mainly use data from SMSS DR2 (O19) and {\it Gaia} EDR3 ({\it Gaia} Collaboration et al. 2020).

The SMSS aims to construct a digital image of the entire Southern Hemisphere sky, with a limiting magnitude of $20$--$22$ in the $uvgriz$ bands (Wolf et al. 2018; O19).
The special design of the $uv$ filters provide photometric sensitivity to stellar surface gravity and metallicity (Bessell et al. 2011).
The ability of these two filters to extract surface-gravity and metallicity information has been well-demonstrated by several studies using SMSS DR1.1 (e.g., Casagrande et al. 2019; da Costa et al. 2019; Huang et al. 2019; Chiti et al. 2020).
Compared to SMSS DR1.1, SMSS DR2 has now released photometric data for a substantially larger number of astrophysical sources, including more data from the Shallow Survey and the deeper Main Survey.
The SMSS Shallow Survey of short exposures is primarily aimed at providing a robust calibration reference and photometric measurements for stars as bright as $g \sim 9-10$, while the Main Survey is planned to scan the Southern Hemisphere to $g \sim 22$ by the year 2021.
In total, over 500 million unique astrophysical sources with photometric information are released in SMSS DR2.
Recently, the photometric zero-points of the $uvgr$-bands in SMSS DR2 have been recalibrated by Huang et al. (2021), using a spectroscopy-based stellar-color regression (SCR) method (Yuan et al. 2015a) to identify and correct the reddening- and spatially-dependent offsets in the $uv$ bands.
If not specified otherwise, here we adopt the recalibrated SMSS DR2 photometry by Huang et al. (2021) in the current work.

In addition to SMSS DR2, the photometric and astrometric data from {\it Gaia} EDR3 are used in this work ({\it Gaia} Collaboration et al. 2020).
We perform a cross-match between SMSS DR2 and {\it Gaia} EDR3 sources (O19), and apply the following cuts to the stars in common:
\begin{enumerate}[label=\arabic*)]

\item Good photometric quality from SMSS DR2: $u/v\_ngood \ge 1$, $u/v\_flags \le 3$, $class\_star \ge 0.9$

\item $G$, $G_{\rm BP}$, and $G_{\rm RP}$ photometry available from {\it Gaia} EDR3, and with uncertainties smaller than 0.05\,mag

\item Galactic latitude $|b| \ge 10^{\circ}$

\end{enumerate}
The last cut is made in order to exclude stars in the Galactic plane, since most of those stars have large values of extinction (which is difficult to estimate accurately).
In total, over 39 million stars (hereafter the Main Sample) remain after the above cuts, and are used in the following analysis.

Finally, all magnitudes and colors presented in the following analysis refer to de-reddened values, corrected using reddening values taken from the extinction map of Schlegel et al. (1998; hereafter SFD98)\footnote{The values of $E (B - V)$ provided by the SFD98 map are corrected for a 14\% systematic over-estimate in the map (e.g., Schlafly et al. 2010; Yuan et al. 2013).}.
The reddening coefficients for the colors $u - G_{\rm BP}$, $v - G_{\rm BP}$, and $G_{\rm BP} - G_{\rm RP}$, and the $G$-band magnitude are color-dependent (e.g.,  Niu et al. 2021), due to the broad {\it Gaia} bands.
Detailed values of these coefficients are specified in the Appendix, as predicted by an $R_{V} = 3.1$  Fitzpatrick extinction law (Fitzpatrick 1999), at $E (B - V) = 0.7$, for different types of stellar spectra. 
The reddening coefficients for the SkyMapper filters are taken from Table\,2 of Huang et al. (2021).

\section{Metallicity}

\begin{figure*}
\begin{center}
\includegraphics[scale=0.325,angle=0]{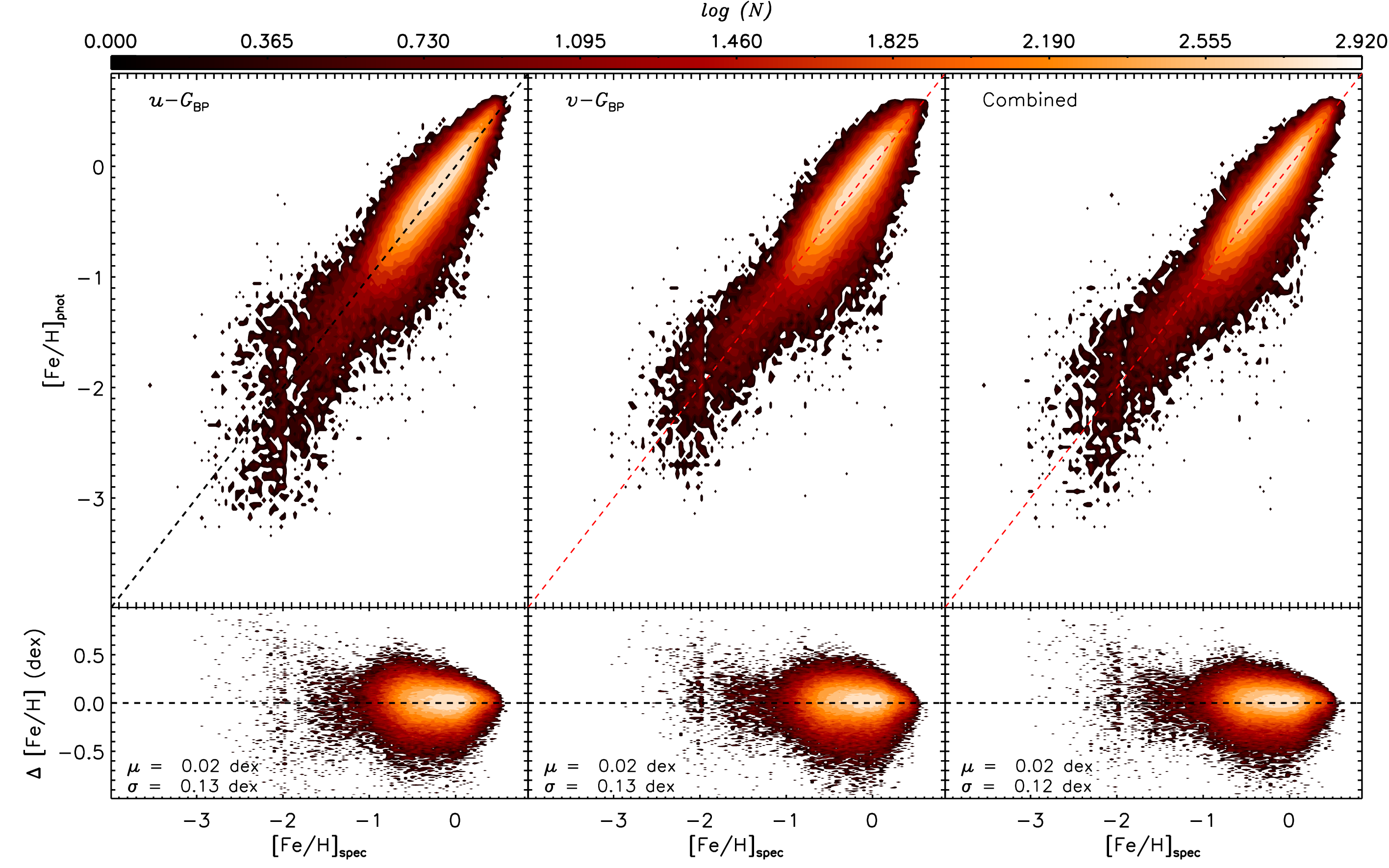}
\caption{ Comparisons between the spectroscopic and photometric metallicities for the training-sample dwarf stars, derived from the colors $u - G_{\rm BP}$ (left panel), $v - G_{\rm BP}$ (middle panel), and the combination of the two colors (right panel).
The lower part of each panel shows the metallicity difference (photometric minus spectroscopic) as a function of the spectroscopic metallicity, with the values of the median and standard deviation of the difference marked in the bottom-left corner.
In each panel, the color-coded contour of the stellar number density on a logarithmic scale is shown.}
\end{center}
\end{figure*}

\begin{figure*}
\begin{center}
\includegraphics[scale=0.325,angle=0]{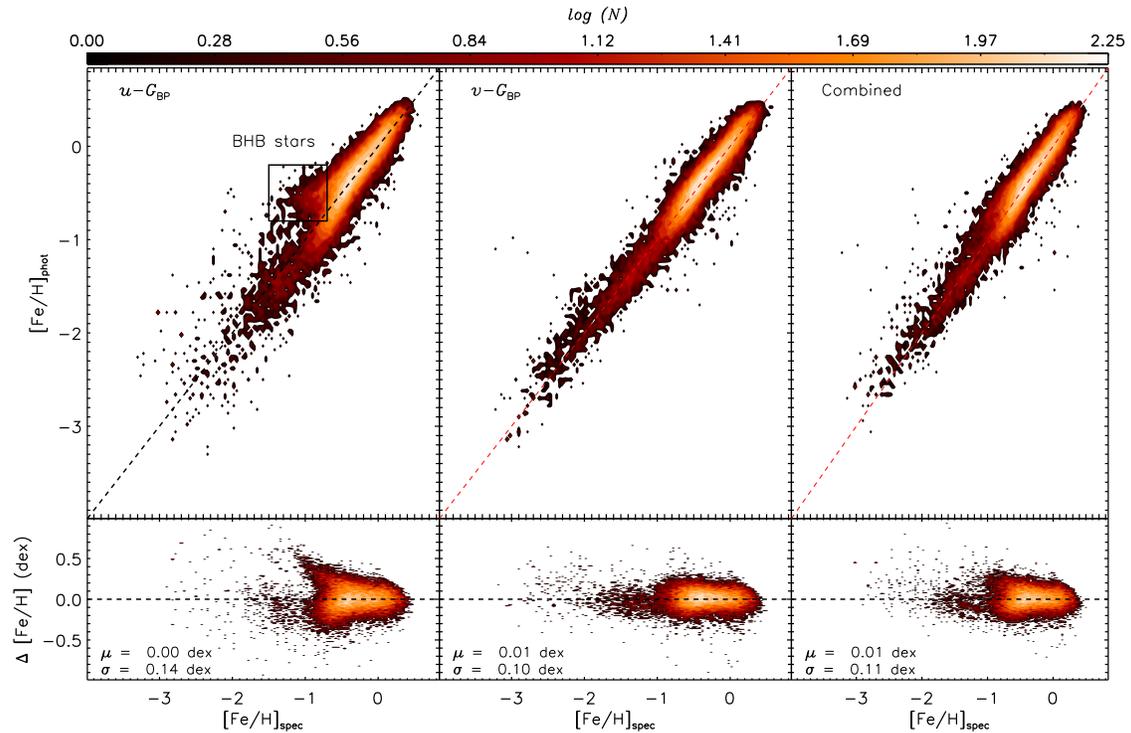}
\caption{Similar to Fig.\,6, but for the training-sample giant stars. The black box in the left panel marks the position where photometric-metallicity estimate deviates from the spectroscopic estimate by larger than expected. This region is dominated by warm blue giants, such as blue horizontal-branch stars.}
\end{center}
\end{figure*}

\begin{figure*}
\begin{center}
\includegraphics[scale=0.395,angle=0]{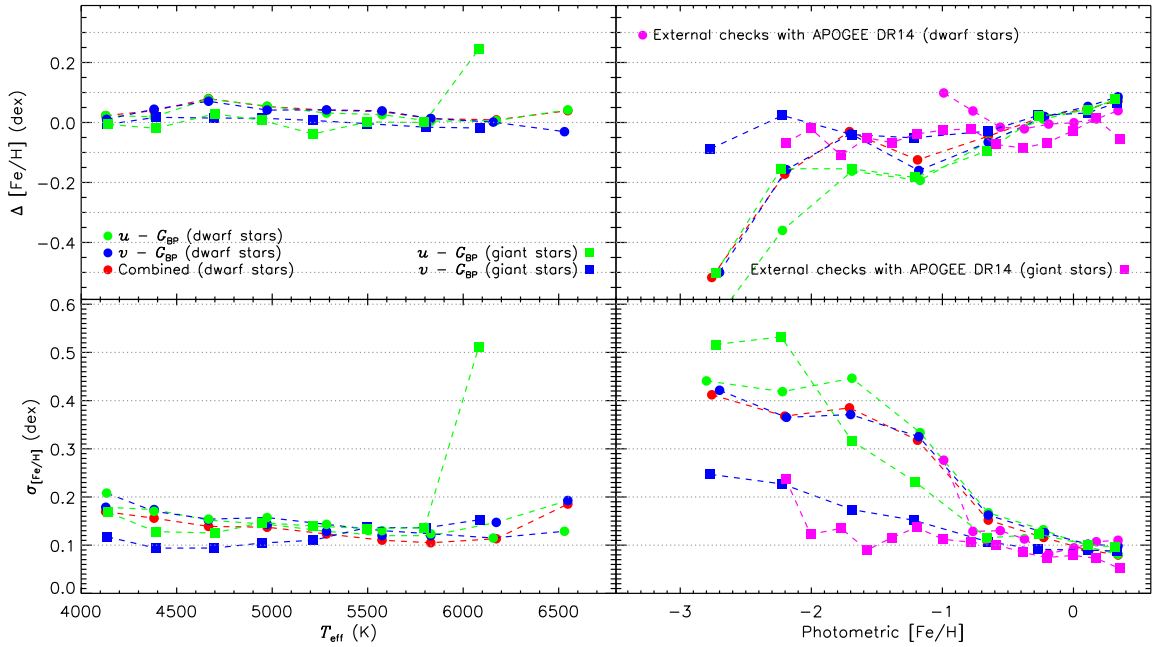}
\caption{Median offsets and standard deviations of the metallicity differences (photometric minus spectroscopic), as a function of effective temperature (left panels) and photometric [Fe/H] (right panels), as calculated from the training sample (dots for dwarf stars and squares for giant stars) and external comparisons with stars from APOGEE DR14 (magenta dots and squares).}
\end{center}
\end{figure*}

 \begin{table*}
\centering
\begin{threeparttable}
\caption{Fit Coefficients for Metallicity and Effective Temperature Estimates for Dwarf and Giant Stars, and Absolute Magnitude Estimates for Main-Sequence Stars}
\begin{tabular}{c|rrr|rrr|r}
\hline
\multirow{2}{*}{Coeff.} & \multicolumn{3}{c}{Dwarf Stars}&\multicolumn{3}{|c}{Giant Stars}&\multicolumn{1}{|c}{\multirow{2}{*}{$M_{G_0}$}\tnote{d}}\\
&$(u - G_{\rm BP})_0$\tnote{a}&$(v - G_{\rm BP})_0$\tnote{b}&$\theta_{\rm eff}$ (K$^{-1}$)\tnote{c}&$(u - G_{\rm BP})_0$\tnote{a}&$(v - G_{\rm BP})_0$\tnote{b}&$\theta_{\rm eff}$ (K$^{-1}$)\tnote{c}&\\
\hline
$a_{0,0}$&$2.859552$&$1.473111$&$0.49815160$&$4.479333$&$0.947087$&$0.51121187$&$6.856251$\\
$a_{0,1}$&$-0.307913$&$-0.198389$&$-0.00952716$&$-0.726008$&$-0.440678$&$-0.03375298$&$-1.140742$\\
$a_{0,2}$&$ 0.021590$&$0.059134$&$-0.00566445$&$0.028400$&$0.049358$&$-0.00657616$&$0.202931$\\
$a_{0,3}$&$0.006480$&$0.003613$&--&$0.007603$&$-0.005236$&--&$-0.035805$\\
$a_{1,0}$&$-7.070455$&$-4.264561$&$0.47924155$&$-11.541715$&$-2.455853$&$0.46326634$&$-9.354844$\\
$a_{1,1}$&$1.112201$&$0.849405$&$-0.01243558$&$1.769734$&$1.247865$&$0.01650766$&$1.594512$\\
$a_{1,2}$&$0.058309$&$-0.019807$&--&$0.055898$&$-0.030162$&--&$-0.421643$\\
$a_{2,0}$&$8.891485$&$6.467822$&$-0.00690521$&$12.715019$&$4.283295$&$0.00153642$&$11.806629$\\
$a_{2,2}$&$-0.444046$&$-0.371660$&--&$-0.701556$&$-0.502618$&--&$-0.773851$\\
$a_{3,0}$&$-2.771184$&$-2.091746$&--&$-3.699729$&$-1.201157$&--&$-3.635873$\\ 
\hline

\hline
\end{tabular}
\begin{tablenotes}
\item[a]$(u - G_{\rm BP})_0 = a_{0,0} + a_{0,1}y + a_{0,2}y^2 + a_{0,3}y^3 + a_{1,0}x + a_{1,1}xy + a_{1,2}xy^2 + a_{2,0}x^2 + a_{2,1}x^2y + a_{3,0}x^3$, where $x$ and $y$ represent $(G_{\rm BP} -  G_{\rm RP})_0$ and [Fe/H], respectively.
\item[b]$(v - G_{\rm BP})_0 =  a_{0,0} + a_{0,1}y + a_{0,2}y^2 + a_{0,3}y^3 + a_{1,0}x + a_{1,1}xy + a_{1,2}xy^2 + a_{2,0}x^2 + a_{2,1}x^2y + a_{3,0}x^3$, where $x$ and $y$ represent $(G_{\rm BP} -  G_{\rm RP})_0$ and [Fe/H], respectively.
\item[c] $\theta_{\rm eff}  =  a_{0,0} + a_{0,1}y + a_{0,2}y^2 + a_{1,0}x +  a_{1,1}xy + a_{2,0}x^2$, where $x$ and $y$ represent $(G_{\rm BP} -  G_{\rm RP})_0$ and [Fe/H], respectively, and $\theta_{\rm eff} = 5040/T_{\rm eff}$.
\item[d]$M_{G_0} =a_{0,0} + a_{0,1}y + a_{0,2}y^2 + a_{0,3}y^3 + a_{1,0}x + a_{1,1}xy + a_{1,2}xy^2 + a_{2,0}x^2 + a_{2,1}x^2y + a_{3,0}x^3$, where $x$ and $y$ represent $(G_{\rm BP} -  G_{\rm RP})_0$ and [Fe/H], respectively.
\end{tablenotes}
\end{threeparttable}
\end{table*}

\begin{figure*}
\begin{center}
\includegraphics[scale=0.35,angle=0]{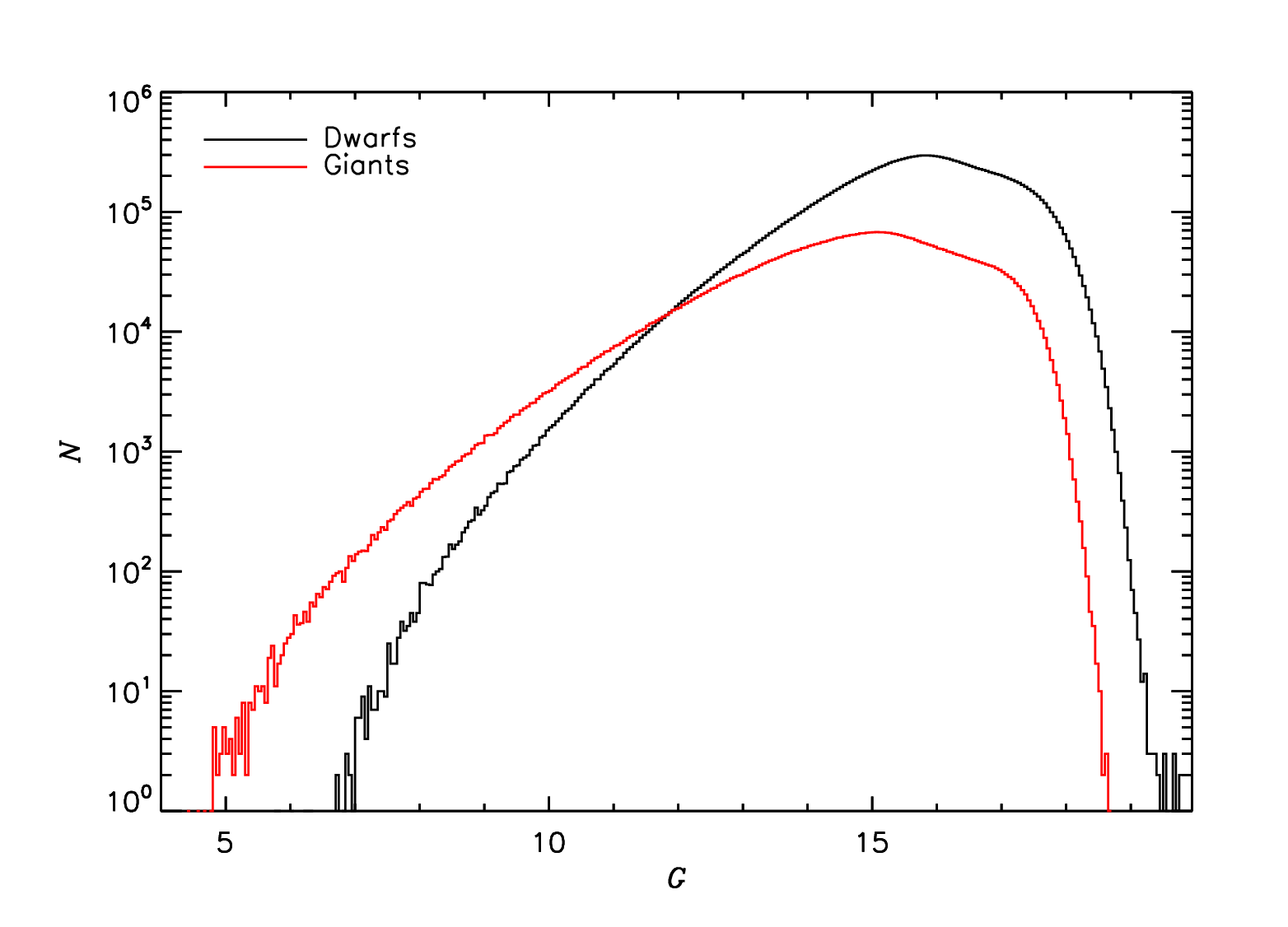}
\includegraphics[scale=0.35,angle=0]{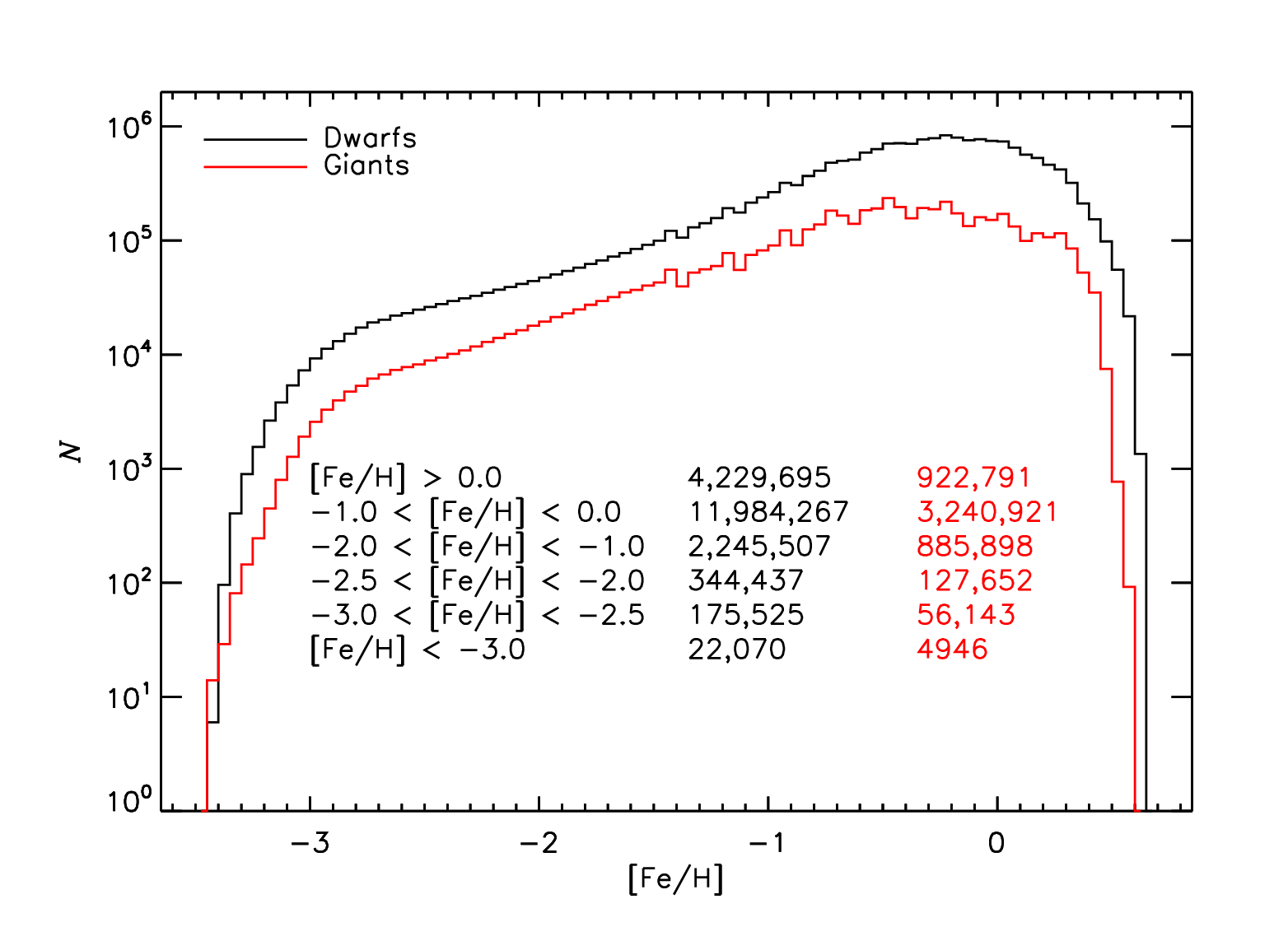}
\caption{{\it Left panel:} Magnitude distributions of dwarf (black line) and giant (red line) stars with photometric metallicity determined from the SMSS DR2  and {\it Gaia} EDR3 color(s).
{\it Right panel:} The photometric-metallicity distributions of dwarf (black line) and giant (red line) stars. The number of dwarf (black) and giant (red) stars for individual metallicity bins are also given.}
\end{center}
\end{figure*}

\begin{figure}
\begin{center}
\includegraphics[scale=0.38,angle=0]{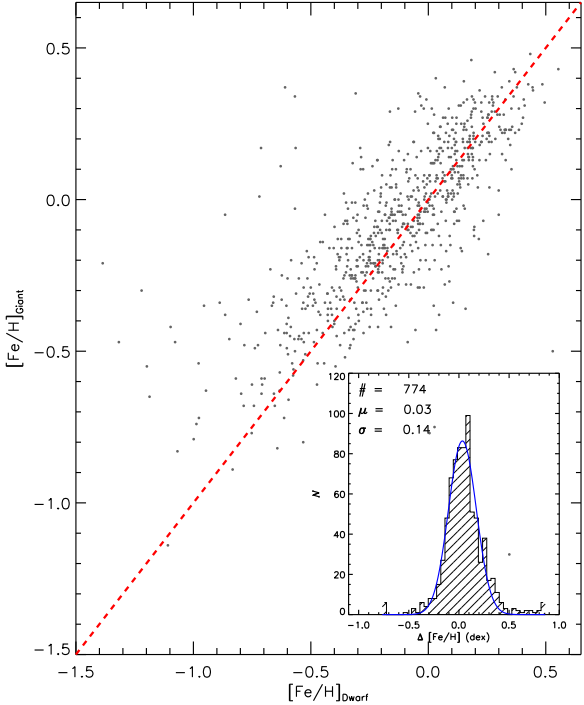}
\caption{Comparison of photometric-metallicity estimates for  dwarf and giant stars in binary systems. 
In the bottom-right corner, the difference in metallicity between giant and dwarf stars is shown. The blue line is a Gaussian fit to the distribution, with the total number of dwarf-giant binary systems, the mean, and dispersion of the Gaussian is marked in the plot.}
\end{center}
\end{figure}

\begin{figure*}
\begin{center}
\includegraphics[scale=0.35,angle=0]{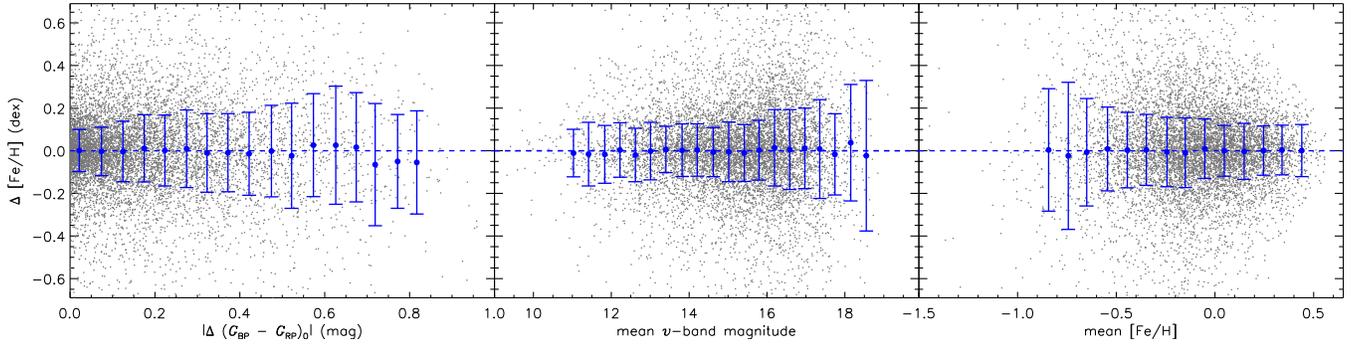}
\caption{Photometric metallicity differences between the two stars in identified binary systems, as a function of absolute color differences of the two stars (left panel), mean $v$ band magnitude (middle panel) and mean [Fe/H] (right panel).
The blue dots and error bars in each panel represent the median and dispersion of the metallicity differences in the individual absolute color difference, mean $v$-band magnitude, and mean [Fe/H]  bins.}
\end{center}
\end{figure*}

\begin{figure*}
\begin{center}
\includegraphics[scale=0.295,angle=0]{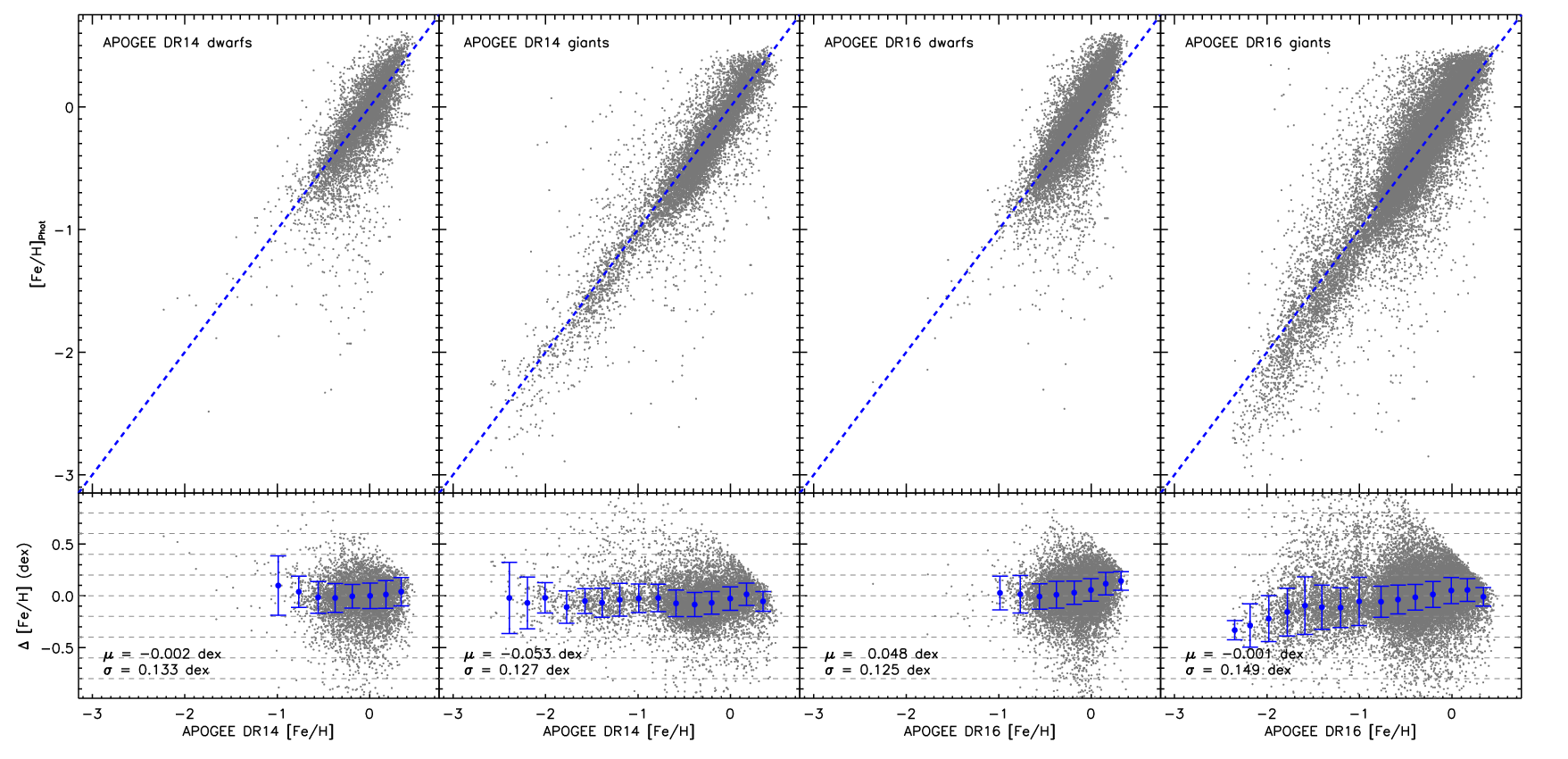}
\caption{Comparisons of photometric-metallicity estimates with those from APOGEE DR14 and DR16.
The differences are shown in the lower part of each panel, with the overall median and standard deviation marked in the bottom-left corner.
The blue dots and error bars in each panel represent the median and dispersion of the metallicity differences in the individual metallicity bins.}
\end{center}
\end{figure*}

\begin{figure*}
\begin{center}
\includegraphics[scale=0.335,angle=0]{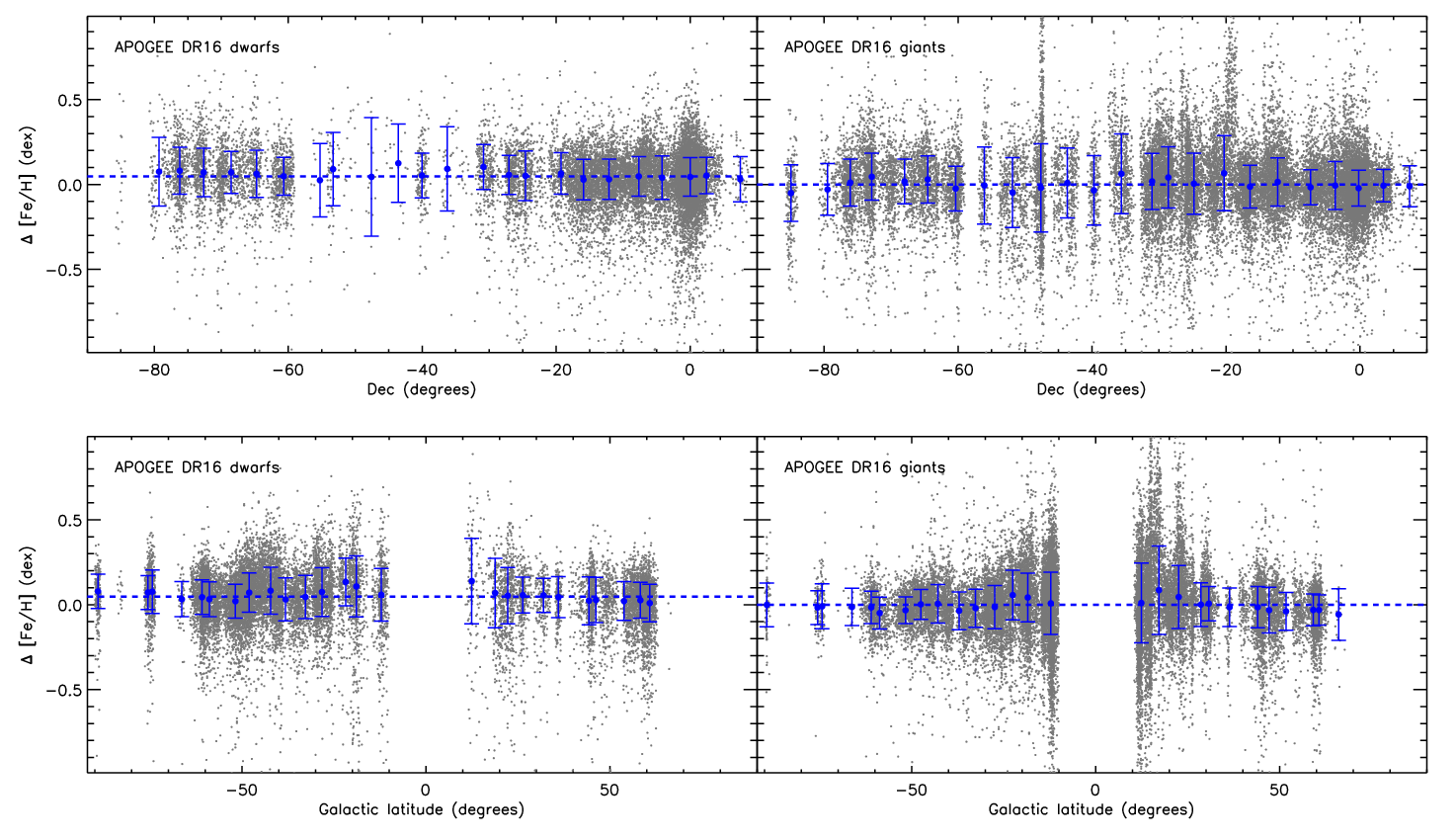}
\caption{Differences between photometric-metallicity estimates and those from APOGEE DR16 for dwarf (left) and giant (right) stars, as a function of declination (top panels) and Galactic latitude (bottom panels).
The blue dots and error bars represent the medians and standard deviations of the differences in the individual declination or Galactic latitude bin.}
\end{center}
\end{figure*}

\begin{figure}
\begin{center}
\includegraphics[scale=0.305,angle=0]{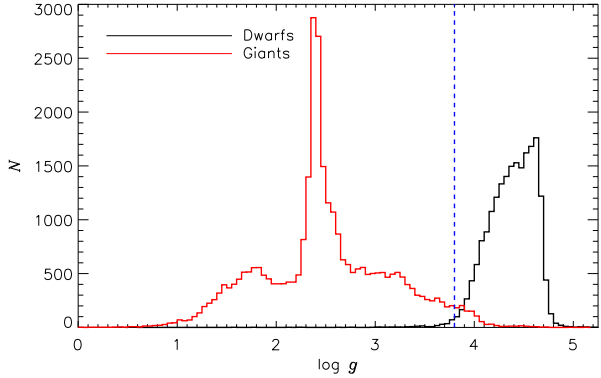}
\caption{Distributions of surface gravity (log\,$g$) given by APOGEE DR16 for the dwarf (black line) and giant (red line) stars, as classified by the cuts defined in Fig.\,3. The dashed blue line represents log\,$g = 3.8$.}
\end{center}
\end{figure}

\begin{figure*}
\begin{center}
\includegraphics[scale=0.27,angle=0]{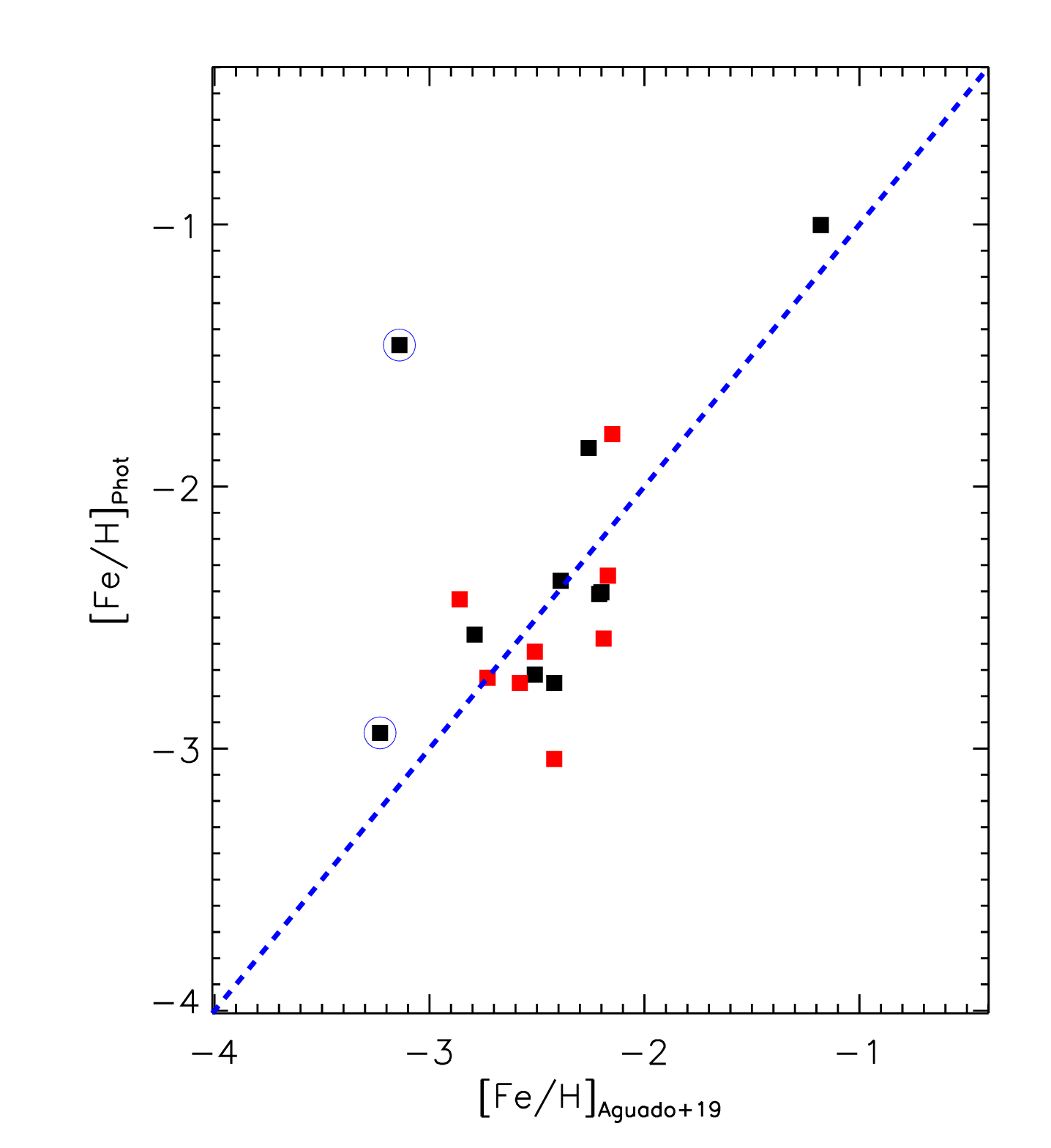}
\includegraphics[scale=0.27,angle=0]{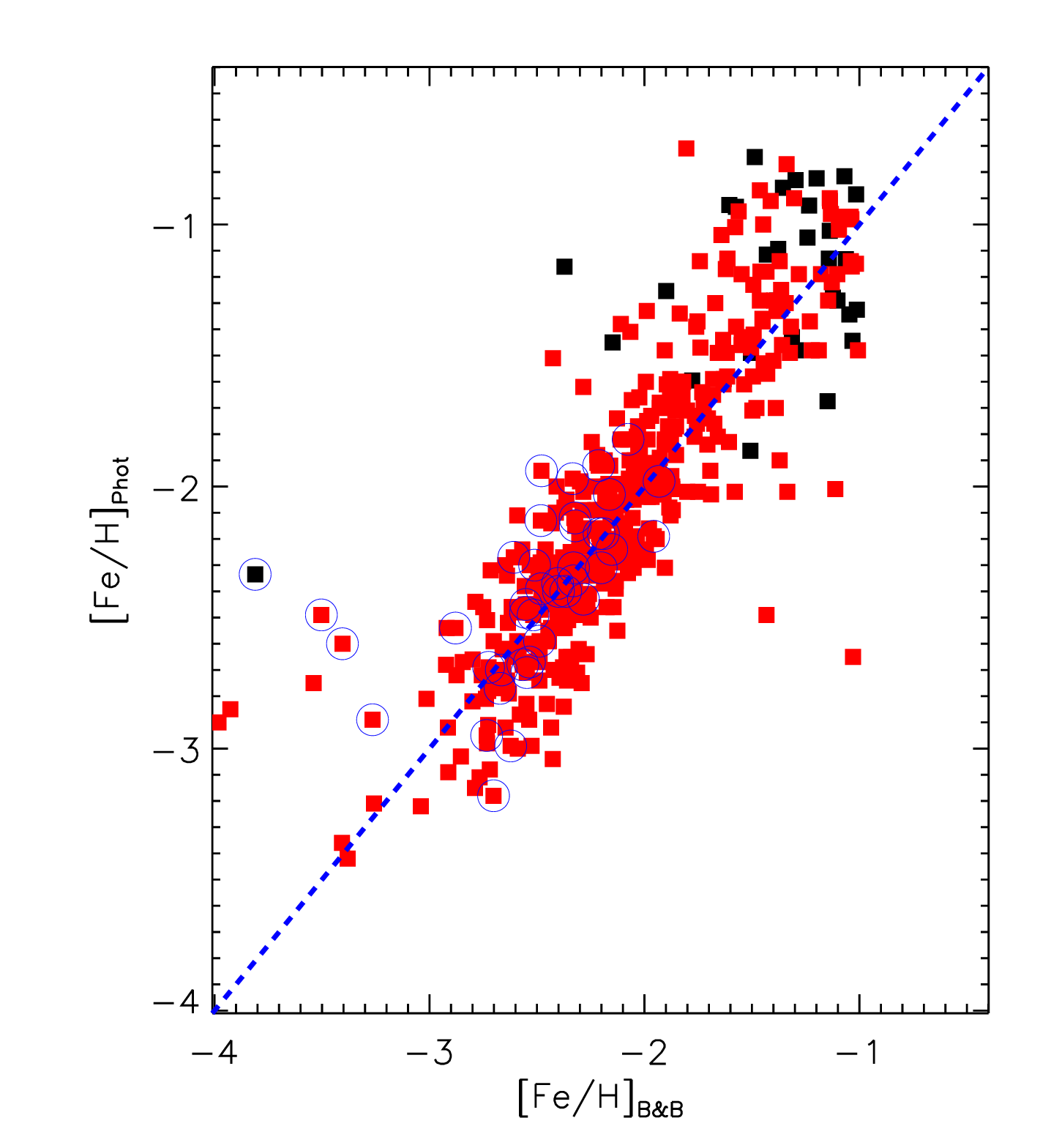}
\includegraphics[scale=0.27,angle=0]{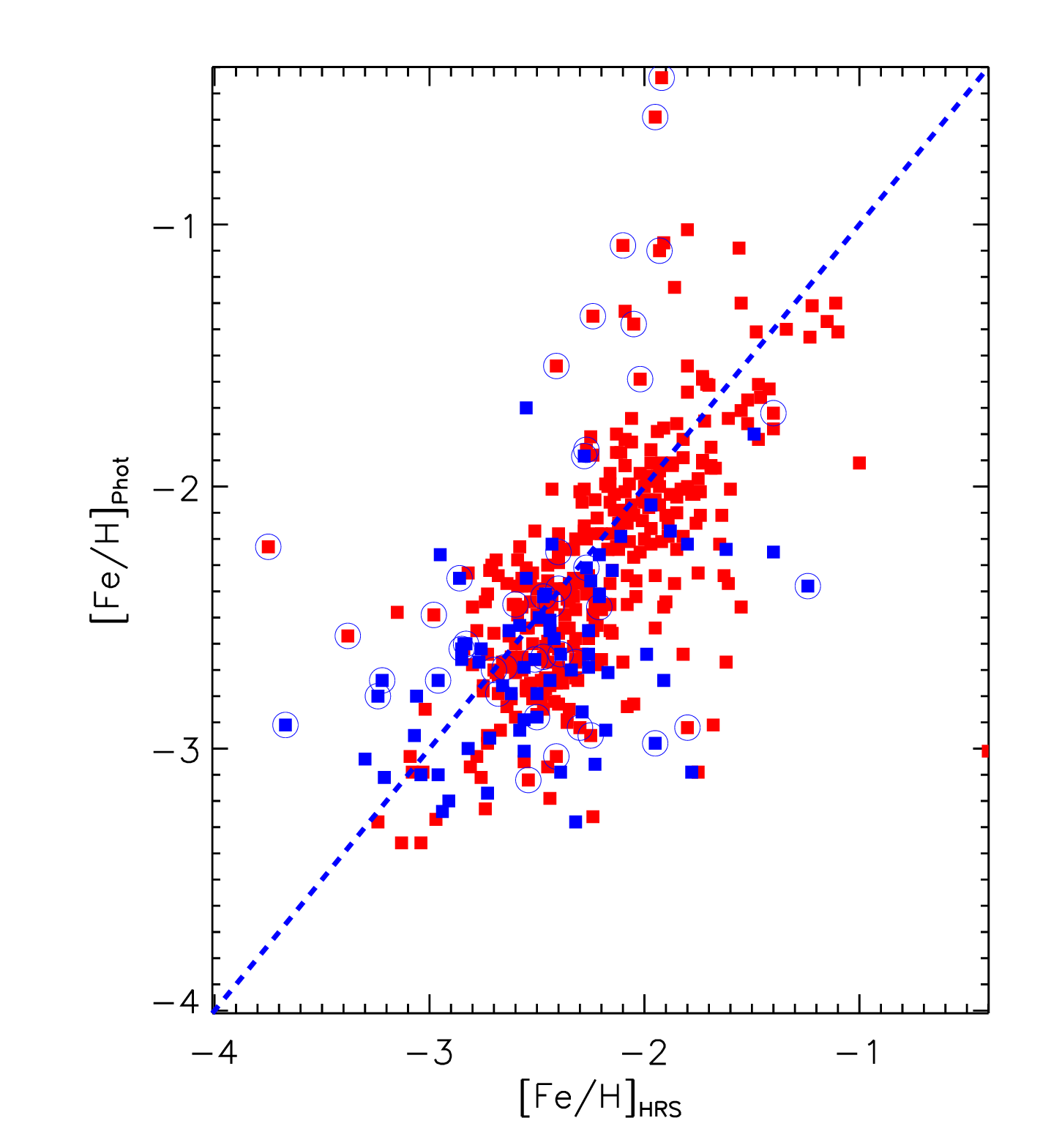}
\caption{{\it Left panel:} Comparison between our photometric and the medium-resolution spectroscopic estimates of metallicity from Aguado et al. (2019). The black and red squares represent dwarf and giant stars, respectively.
The blue circles mark the carbon-enhanced stars ([C/Fe]\,$> +0.6$).
The significant outlier in the top-left corner is a metal-poor carbon-enhanced star ([C/Fe] = +1.08]).
{\it Middle panel:} Comparison between our photometric and the medium-resolution spectroscopic estimates of metallicity from the Best \& Brightest Survey (B\&B) as reported by  Limberg et al. (2021). The black and red squares represent dwarf and giant stars, respectively.
The blue circles mark the carbon-enhanced stars ([C/Fe]\,$> +0.6$).
{\it Right panel:} Comparison between our photometric and high-resolution spectroscopic estimates of metallicity from the $R$-Process Alliance (RPA) sample (red squares: Hansen et al. 2018; Sakari et al. 2018; Ezzeddine et al. 2020; Holmbeck et al. 2020) and the SkyMapper EMP star candidate sample (blue squares: Jacobson et al. 2015; Marino et al. 2019). Note that the latter determinations include NLTE corrections. The blue circles mark the carbon-enhanced stars ([C/Fe]\,$> +0.6$). }
\end{center}
\end{figure*}

\begin{figure*}
\begin{center}
\includegraphics[scale=0.395,angle=0]{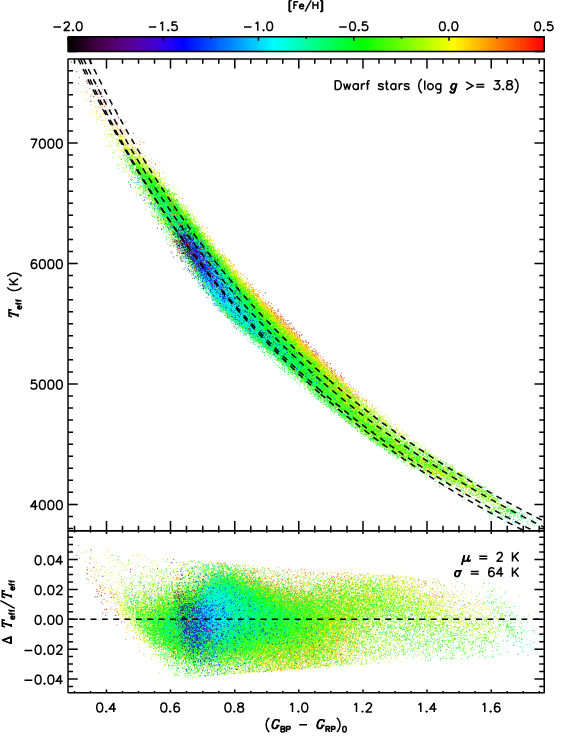}
\includegraphics[scale=0.395,angle=0]{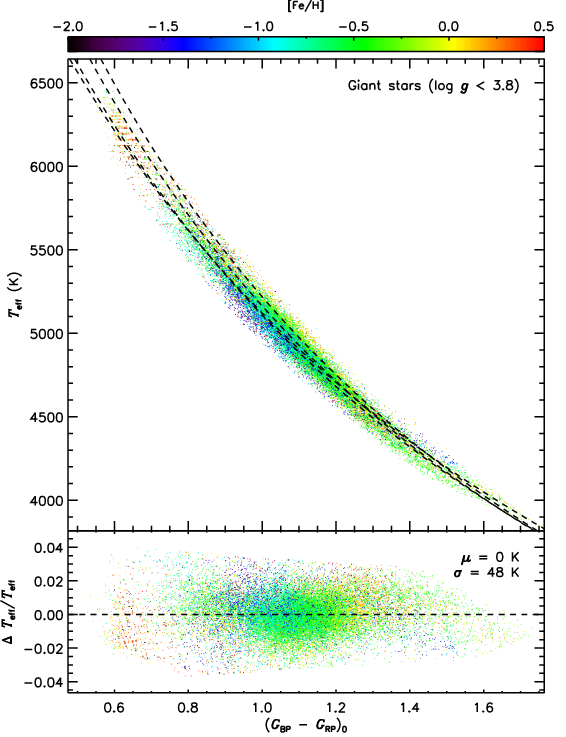}
\caption{$T_{\rm eff}$, as a function of color $(G_{\rm BP} - G_{\rm RP})_0$, for the dwarf (left panel) and giant (right panel) training set color-coded by metallicity, as shown in the top color bar.
The black dashed lines represent our best fits for different values of [Fe/H], as described by Equation\,5. 
From top to bottom, the values of [Fe/H] are $+$0.5, 0.0, $-1.0$ and $-2.0$, respectively.
The lower part of each panel shows the relative fit residual $(T_{\rm eff}^{\rm fit} - T_{\rm eff}^{\rm LM})/T_{\rm eff}^{\rm LM}$, as a function of color $(G_{\rm BP} - G_{\rm RP})_0$, with the values of median and standard deviation of the residual marked in the top-right corner.}
\end{center}
\end{figure*}

\begin{figure*}
\begin{center}
\includegraphics[scale=0.4,angle=0]{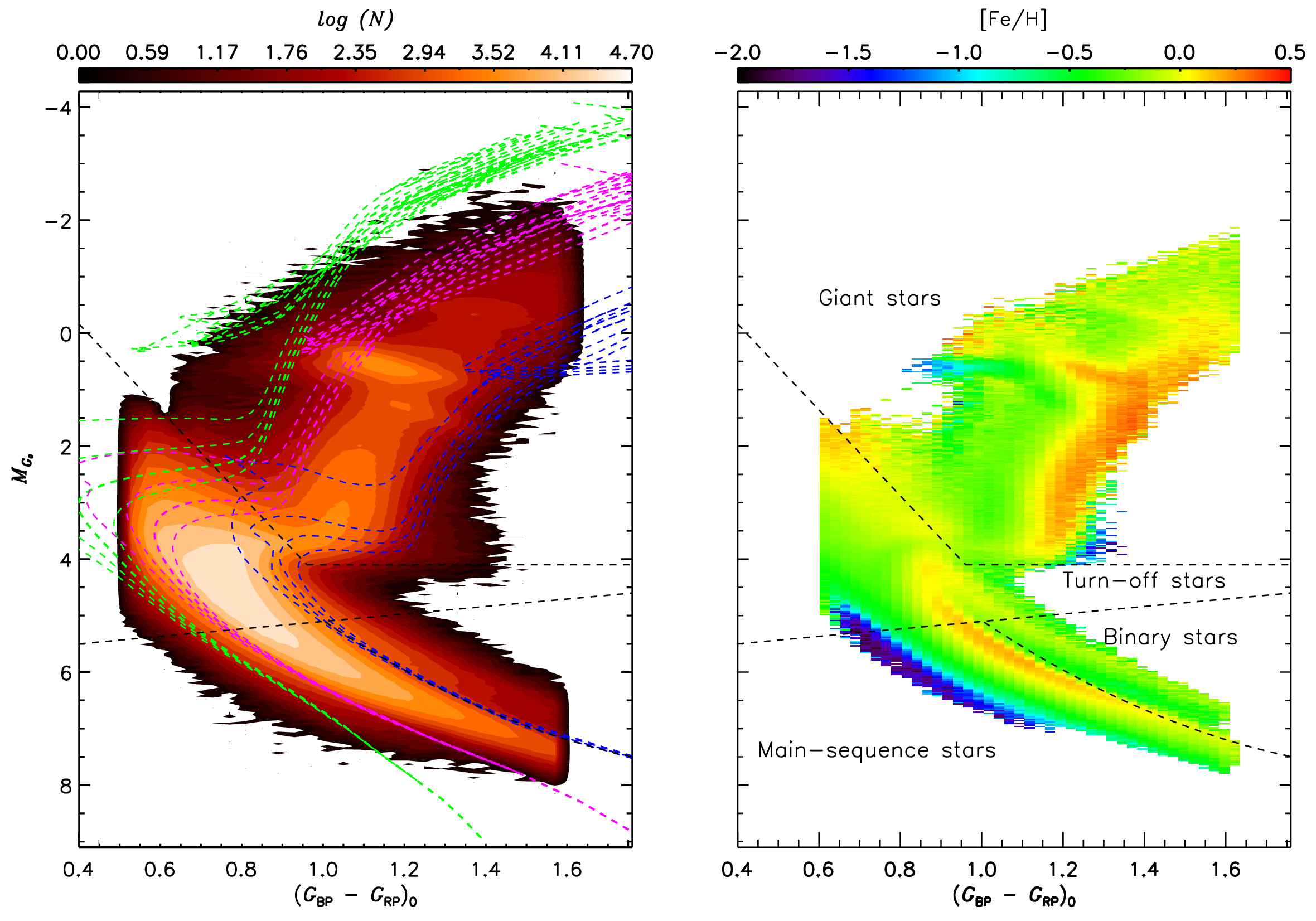}
\caption{{\it Left panel:} Color-coded stellar number-density distribution (on a logarithmic scale) in the $M_{G_0}$ versus $(G_{\rm BP} - G_{\rm RP})_0$ plane. Blue, magenta, and green dashed lines represent stellar isochrones from PARSEC (Bressan et al. 2012; Marigo et al. 2017) with [M/H]\,=\,$+0.50$, $-0.75$ and $-2.00$, with ages from 3 to 9\,Gyr in steps of 2\,Gyr (from left to right). 
The upper black dashed lines (defined in Fig.\,3) separate dwarf and giant stars.
The middle dashed line marks the turn-off stars (above this line).
The lower dashed line separates main-sequence (left part) and binary stars (right part).
{\it Right panel:} Similar to the left panel but color-coded with photometric [Fe/H], as indicated by the top color bar.}
\end{center}
\end{figure*}

 \subsection{Training Set}
 To derive photometric-metallicity estimates from the SkyMapper and {\it Gaia} colors, we require a training sample with reliable spectroscopic-metallicity estimates, as well as good-quality stellar colors.
To accomplish this, we assembled a number of spectroscopic catalogs, including  the Stellar Abundances for Galactic Archaeology (SAGA) Database (Suda et al. 2008), the PASTEL catalog (Soubiran et al. 2016), LAMOST DR7\footnote{http://dr7.lamost.org/}(Luo et al. 2015), and SDSS DR16 (Ahumada et al. 2020)\footnote{Note that this sample includes stars from the SDSS Legacy survey (York et al. 2000), SEGUE (Yanny et al. 2009), SEGUE-2 (Rockosi et al. 2021), BOSS (Dawson et al. 2013), and eBOSS (Dawson et al. 2016).  For simplicity, we refer to this entire set as SDSS/SEGUE, since these sources are the dominant contributors.}.
The atmospheric parameters in LAMOST and SDSS/SEGUE are determined by the LAMOST stellar parameter pipeline (LASP; Wu et al. 2014) and the SEGUE stellar parameter pipeline (SSPP; Allende Prieto et al. 2008; Lee et al. 2008a,b), respectively.
The LASP parameter determination pipeline only provides reliable estimates for stars with {[Fe/H]\,$> -2.0$, and the uncertainties of derived [Fe/H] can be quite large for stars with [Fe/H]\,$< -2.0$.  We have thus redetermined metallicity estimates for LAMOST stars with [Fe/H]$< -1.8$ using a custom version of the SSPP (LSSPP; Lee et al. 2015), and visually inspected (by Beers) to reject clearly problematic stars, including cool white dwarfs, hot B-type sub-dwarf stars, composite spectra, emission line objects, and those with spectroscopic defects that precluded accurate metallicity estimates.
For stars with metallicity estimates available from both LASP and the LSSPP, we adopted the latter.  Similarly, redetermined metallicities for SDSS/SEGUE stars with [Fe/H] $< -1.8$, based on a recent recalibration of the SSPP (Lee et al., in preparation), and visually inspected for problematic stars, are obtained.  
The metallicity estimates for LAMOST and SDSS/SEGUE stars based on the above approaches are reliable down to [Fe/H] $\sim -4.0$, where limitations due to the presence of interstellar contamination of the primary metallicity estimator (Ca {\sc II} K) or from intrinsic enhanced carbon become significant. 

Finally, we choose the metallicity of the SMSS stars that appear in SAGA and/or PASTEL to establish a standard scale based on high-resolution spectroscopy (HRS).
We note that this standard scale could also suffer from potential systematic biases, since metallicities in the SAGA and PASTEL databases are simply bibliographical compilations of the measurements from different groups, determined with different methods, atmospheric models, and stellar spectral properties.
To quantitatively evaluate this systematic biases, we show the standard deviations of [Fe/H] as a function of [Fe/H] with number of measurements greater than 4 for over 2000 stars in the PASTEL database in Fig.\,1.
The plot shows that the metallicity difference among different groups is typically about 0.00--0.05\,dex for stars with [Fe/H]\,$> -2.0$ and 0.05--0.10\,dex for stars with [Fe/H]\,$\leq -2.0$, respectively.
The relatively larger deviations for metal-poor stars ([Fe/H]\,$\leq -2.0$) is contributed in part by metallicity determinations based on different adopted temperature scales and/or obtained with/without considering non-local thermodynamic equilibrium (NLTE) effects by different groups.

A comparison of metallicity determinations based on low-resolution spectroscopy for the LAMOST stars in common with those from the SAGA and PASTEL databases is shown in the left panel of Fig.\,2.
Generally, the metallicity of the two samples are consistent with each other, with deviations visible at the lowest metallicities.  To tie the metallicity scale of LAMOST to that of the HRS sample, a linear fit:

\begin{equation}
{\rm [Fe/H]_{HRS}} = 0.019 + 1.091 \times {\rm [Fe/H]_{\rm LRS}}\text{,}
\end{equation}

\noindent is obtained from the comparison.

To further calibrate the metallicity scale based on the SDSS/SEGUE stars, a comparison between metallicity estimates for the SDSS/SEGUE and the calibrated LAMOST stars (metallicity scale corrected to the HRS sample using Equation\,1) in common is shown in the right panel of Fig.\,2.

To tie the metallicity scale of SDSS/SEGUE to that of the calibrated LAMOST sample, a linear fit:

\begin{equation}
{\rm [Fe/H]_{LRS}^{calibrated}} =  0.098 + 1.131\times{\rm [Fe/H]}_{\rm SDSS/SEGUE}\text{,}
\end{equation}

\noindent is obtained from the comparison.
Based on the above approach, the metallicity scales for the LAMOST and SDSS/SEGUE stars are all calibrated to that of the HRS sample.
We note that few outliers in both plots are possibly due to unrecognized issues in the metallicity estimates from the low/medium-resolution spectra. However, the above relations are not affected by those outliers, which are clipped in our fitting process.
 
 The combined spectroscopic sample of these stars are then cross-matched with the Main Sample defined above. In total, over 190,000 stars are found in common.
 To achieve a training set of high quality, the following cuts are applied to these stars:
 
 \begin{enumerate}[label=\arabic*)]

\item The stars must have Galactic latitude $|b| \ge 20^{\circ}$ and $E (B - V) \leq 0.07$ to minimize uncertainties due to reddening corrections

\item The LAMOST and SDSS/SEGUE stars must have spectral signal-to-noise ratio (SNR) greater than 20, and all stars must have temperatures in the range 4000 $\leq T_{\rm eff}$\,${\rm (K)} \leq$ 6800 (i.e., typical FGK-type stars) to ensure high-precision metallicity estimates

\item The photometric uncertainties in the SkyMapper $uv$ and {\it Gaia} $G_{\rm BP}G_{\rm RP}G$ bands must be smaller than 0.035\,mag

\item The stars must have $Gaia$ parallax measurement uncertainties smaller than 30\%

\end{enumerate}

With the above choices, 172,501 stars are finally selected to construct the training set.
By adopting distances from Bailer-Jones et al. (2021), the absolute magnitude in the $G$ band is obtained for these stars; the resulting Hertzsprung–Russell (H-R) diagram of the training set is shown in the left panel of Fig.\,3.
By using empirical cuts: $M_{G_0} = -3.20 + 7.60\cdot (G_{\rm BP} - G_{\rm RP})_0$ or $M_{G_0} = 4.1$, shown in this diagram, the training stars are further divided into dwarfs and giants.
The metallicity distributions of the dwarf and giant stars in the training set are shown in the right panel of Fig.\,3.
 
 \subsection{Metallicity-Dependent Stellar Loci}
Based on (recalibrated) SDSS/Stripe 82 photometry, Yuan et al. (2015b) investigated the intrinsic widths of the SDSS stellar loci after considering the effects of metallicity.  They found that the intrinsic widths of the metallicity-dependent loci are at most a few mmags, if not zero.
Based on the above fact, Yuan et al. (2015c) developed a method to determine photometric metallicities by fitting the dereddened SDSS colors to the empirically determined metallicity-dependent stellar loci. With 1\% photometry, the method achieved a precision of 0.05, 0.12, and 0.18 dex at metallicities of [Fe/H] = 0.0, $-1.0$, and $-2.0$, respectively.
In this work, we use a similar technique for metallicity determinations. 
 
The metallicity-dependent stellar loci of $(u - G_{\rm BP})_0$ and $(v - G_{\rm BP})_0$ versus $(G_{\rm BP} - G_{\rm RP})_0$ are shown in Figs.\,4 and 5 for dwarf and giant stars, respectively, using the aforementioned training set. 
The plots clearly show sequences of different metallicities ranging from [Fe/H] = $-3.5$ to [Fe/H] = $+0.5$, as both colors $(u - G_{\rm BP})_0$ and $(v - G_{\rm BP})_0$ change with the $(G_{\rm BP} - G_{\rm RP})_0$ for typical FGK-type stars.
To quantitatively describe these stellar loci, third-order 2-D polynomials (with 10 free parameters, including cross terms) are adopted to fit the $(u - G_{\rm BP})_0$ and $(v - G_{\rm BP})_0$ colors, as a function of $(G_{\rm BP} - G_{\rm RP})_0$ and [Fe/H], respectively, for dwarf and giant stars:
 \begin{equation}
 \begin{split}
 (u/v - G_{\rm BP})_0 & =  a_{0,0} + a_{0,1}y + a_{0,2}y^2 + a_{0,3}y^3 + a_{1,0}x + \\
 & a_{1,1}xy + a_{1,2}xy^2 + a_{2,0}x^2 + a_{2,1}x^2y + a_{3,0}x^3\text{,}
 \end{split}
 \end{equation}
where $x$ and $y$ represent $(G_{\rm BP} - G_{\rm RP})_0$ and [Fe/H], respectively.
Three-sigma clipping is applied in the fitting process.
The resultant fit coefficients are listed in Table\,2.
 
 \subsection{Metallicity Determinations} 
 
 Using the empirical stellar loci defined above, we adopt the maximum-likelihood approach to derive photometric-metallicity estimates for our program sample. 
 For a given star, the likelihood is defined as:
 \begin{equation}
 L_c = \frac{1}{\sqrt{2\pi}\sigma_{c_{\rm obs}}}\exp{\frac{-(c_{\rm obs} - c_{\rm pred})^2}{2\sigma_{c_{\rm obs}}^2}},
 \end{equation}
 where $c_{\rm obs} = \{(u - G_{\rm BP})_0, (v - G_{\rm BP})_0\}$ are assumed to be independent Gaussian observables.
 The $c_{\rm pred}$ parameter represents the same colors and is a function of $(G_{\rm BP} - G_{\rm RP})_0$ and [Fe/H], which can be predicted from our metallicity-dependent stellar loci (i.e., Equation\,3).
 The value of [Fe/H] is varied from $-3.5$ to $+0.8$, with steps of 0.01\,dex, when predicting $(u/v - G_{\rm BP})_0$.
 
We also define applicability ranges of the current method: (1) It is reliable for limited color ranges in $(G_{\rm BP} - G_{\rm RP})_0$ (roughly 0.56 to 1.61 for dwarf stars and 0.67 to 1.62 for giant stars; see the dotted lines in Figs.\,4 and 5); (2) The applicable upper and lower metallicity limits in color $(u/v - G_{\rm BP})_0$ are defined for individual $(G_{\rm BP} - G_{\rm RP})_0$  bins with steps of 0.05\,mag, using the training sets for dwarf and giant stars, respectively. By combining the applicability ranges and the likelihood function, the probability distribution function (PDF) of [Fe/H] can be separately obtained for each star, either from $(u - G_{\rm BP})_0$ or $(v - G_{\rm BP})_0$.
The photometric metallicity (median value) and its uncertainty (half of the 68\% interval) are then deduced from the resultant PDF.
 
As an internal test, we first determine photometric-metallicity estimates for the training set by the above method. The results are shown in Figs.\,6 and 7.
For dwarf stars, the derived photometric metallicities, either by $(u - G_{\rm BP})_0$ or $(v - G_{\rm BP})_0$, agree with the the spectroscopic values quite well, without obvious offsets or trends for [Fe/H]\,$> -2.0$ (see the top-right panel of Fig.\,8); the resulting overall scatters are around 0.13\,dex.
By combining the estimates from the two colors (using an error-weighted mean), the overall scatter can be further reduced to 0.12\,dex, similar to (or better than) the uncertainty reported for low/medium-resolution spectroscopy. 
We note that a modest offset of around (0.1 -- 0.3\,dex) is found for dwarfs with [Fe/H]\,$< -2.0$ (see the top-right panel of Fig.\,8).
Generally, the scatter revealed by the internal test is a strong function of [Fe/H], with $\sigma_{\rm [Fe/H]}$ of 0.10--0.20\,dex for metal-rich stars ([Fe/H]\,$\ge -1.0$) and $\sigma_{\rm [Fe/H]}$ of 0.20--0.40\,dex for metal-poor stars ([Fe/H]\,$< -1.0$), respectively (see the bottom-right panel of Fig.\,8).
The photometric-metallicity estimates do not exhibit any significant trends with stellar effective temperature (see the bottom-left panel of Fig.\,8).
For giant stars, the photometric-metallicity estimates yielded by the color $(v - G_{\rm BP})_0$ are in excellent agreement with the spectroscopic values, with a overall scatter of only 0.10\,dex (and smaller offsets for [Fe/H] $< -2.0$, compared to dwarfs; see the top-right panel of Fig.\,8). 
The internal precision for the photometric-metallicity estimates derived from the color $(v - G_{\rm BP})_0$ is 0.05-0.10\,dex for stars with [Fe/H]\,$\leq -1.0$, 0.10-0.20\,dex for stars with $-1.0 <$\,[Fe/H]\,$\leq -2.0$ and 0.20-0.25\,dex for stars with [Fe/H]\,$< - 2.0$, respectively (see the bottom-right panel of Fig.\,8).
The performance of the color $(u - G_{\rm BP})_0$ is moderately worse, with significant systematics (see Figs.\,7 and 8), due to the lower sensitivity of  $(u - G_{\rm BP})_0$ on metallicity, and the sensitivity of the $u$-band to surface gravity, most evident for warmer giant stars (e.g., field blue horizontal-branch stars; see the top-left panel of Fig.\,8).
 
According to the above internal test, the final photometric-metallicity of a dwarf star is obtained by the combined estimate, if both $(u - G_{\rm BP})_0$ and $(v - G_{\rm BP})_0$ colors are available, or from a single estimate, either from $(u - G_{\rm BP})_0$ or $(v - G_{\rm BP})_0$, depending on which color is available.
The final photometric-metallicity estimate for a giant star is given by the estimate from the color $(v - G_{\rm BP})_0$, or the color $(u - G_{\rm BP})_0$ if the former is not available.
In this manner, photometric-metallicity estimates are obtained for over 24 million program stars (over 19 million dwarf stars and 5 million giant stars; hereafter the Final Sample) from the Main Sample (see Section\,2). Detailed information on the sample of stars with estimated metallicities is summarized in Table\,3.
The magnitude distributions of dwarf and giant stars with metallicity estimates are shown in the left panel of Fig.\,9.
The distributions show two clear turning points (with one at $G \sim 15.0$ and another at $G \sim 17.5$), resulting from the observations of the Shallow and Main SMSS surveys.
The metallicity distributions for dwarf and giant stars are presented in the right panel of Fig.\,9.  
We note that photometric-metallicity estimates are obtained for unprecedented numbers of stars, including a total of over three million metal-poor (MP; [Fe/H] $< -1.0$) stars, over half a million very metal-poor (VMP; [Fe/H] $< -2.0)$ stars, and over 25,000 extremely metal-poor (EMP; [Fe/H] $< -3.0$) stars.
Finally, the metallicity distributions (after correcting for potential selection effects) will provide vital clues to the formation and evolution of the Galactic stellar halo.

The final uncertainties of the estimated photometric metallicity is given by $\sqrt{\sigma_{\rm m}^2 + \sigma_{\rm r}^2}$, where $\sigma_{\rm m}$ is the error determined by the internal tests (see Figs. 6, 7 and 8), and $\sigma_{\rm r}$ is the random error given by the likelihood as defined in Equation\,4.
 
 \subsection{Validation with Gaia Wide Binaries}

Wide binaries are believed to be born at the same time and location with the same metallicity, and thus can be used to examine our photometric-metallicity measurements.
To accomplish this, we cross-match the wide binary sample (Tian et al. 2020) constructed from {\it Gaia} DR2 (Gaia Collaboration et al. 2018) to the Final Sample, and find 8,795 wide binaries with separations less than 20,000 AU.
Among those wide binaries, 774 are dwarf-giant systems (with classifications given in the current work).
A comparison of the photometric-metallicity estimates between the dwarf and giant stars in these binary systems is shown in Fig.\,10.  From inspection, our photometric-metallicity estimates for dwarf stars agree with those for giant stars very well, with a very small  offset of around 0.03\,dex and a standard deviation of 0.14\,dex, indicating a typical measurement error around 0.10\,dex for both dwarf and giant stars.  
There does appear to exist a discrepancy between these estimates for metallicities below [Fe/H] $= -1.0$, in the sense that the giants are assigned higher metallicities than the dwarfs.
In addition, a few significant outliers are seen in the plot; they are possibly non-physical binary contaminators in the sample.

The metallicity differences between the two stars in these binary systems, as functions of  absolute color difference of the two stars in the binary system, mean $v$-band magnitude, and mean [Fe/H] are shown in Fig.\,11.
Generally, no systematic patterns of the differences are found in those spaces (i.e.,  absolute color difference, mean $v$-band magnitude, and mean [Fe/H]).
Typically, the scatter of the difference increases with absolute color difference and mean $v$-band magnitude, and decrease with mean [Fe/H].
By assuming the scatter of difference is equally contributed by both measurement errors, the typical measurement uncertainty is 0.10\,dex for stars with $v < 16$ and around 0.15\,dex for stars with $v > 16$.

 \begin{figure*}
\begin{center}
\includegraphics[scale=0.4,angle=0]{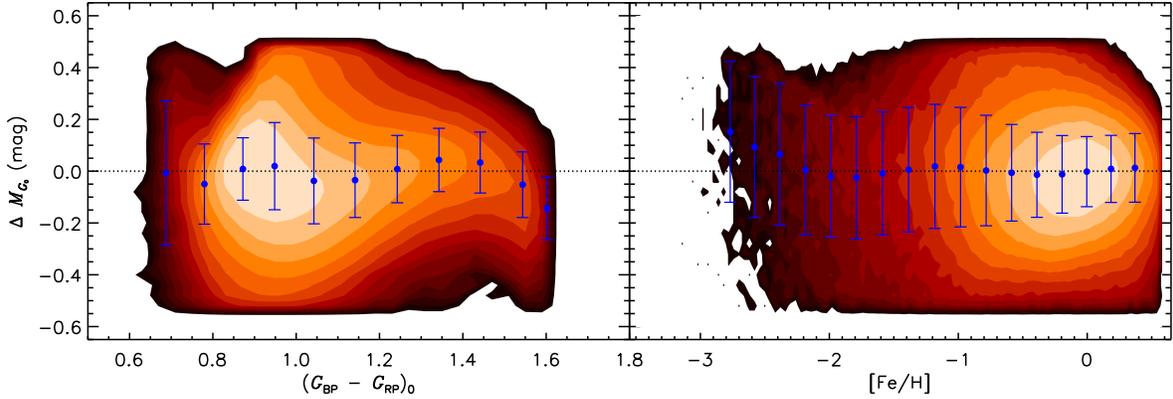}
\caption{The fitting residuals of absolute magnitude $\Delta M_{G_0}$, as a function of stellar color $(G_{\rm BP} - G_{\rm RP})_0$ (left panel) and photometric [Fe/H] (right panel).
The blue dots and error bars represent the medians and standard deviations of the differences in the individual $(G_{\rm BP} - G_{\rm RP})_0$ or [Fe/H] bin.}
\end{center}
\end{figure*}

\begin{figure}
\begin{center}
\includegraphics[scale=0.325,angle=0]{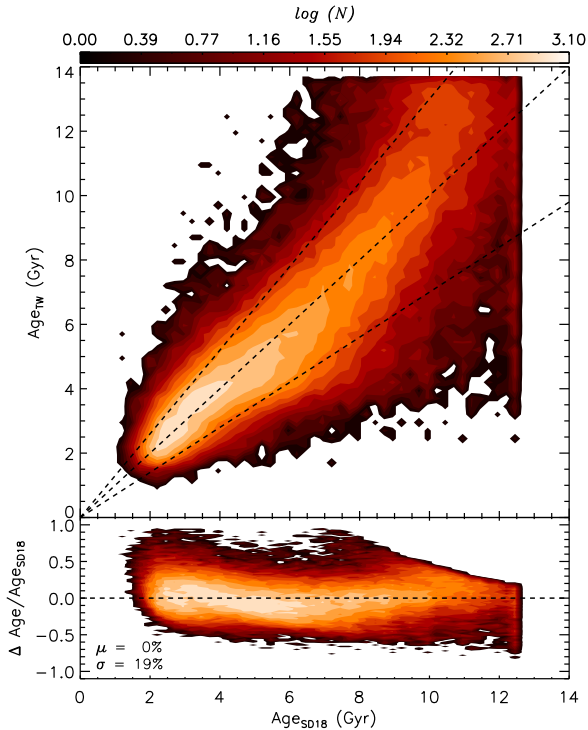}
\caption{Comparisons of stellar-age estimates between this work and SD18 for over 120,000 main-sequence turn-off stars in common. 
The three dashed lines mark Age$_{\rm TW} =$ $1.3$Age$_{\rm SD18}$,  Age$_{\rm TW} =$ Age$_{\rm SD18}$ and Age$_{\rm TW} =$ $0.7$Age$_{\rm SD18}$, respectively.
The lower panel shows the relative age difference (this work minus SD18), as a function of SD18 age, with the values of the median and standard deviation of the relative age difference marked in the bottom-left corner. The color-coded contour of the stellar number density on a logarithmic scale is shown.}
\end{center}
\end{figure}

\begin{figure*}
\begin{center}
\includegraphics[scale=0.325,angle=0]{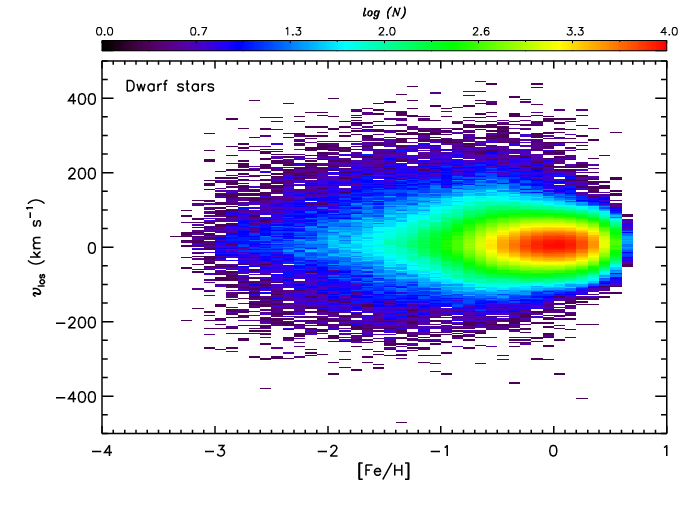}
\includegraphics[scale=0.325,angle=0]{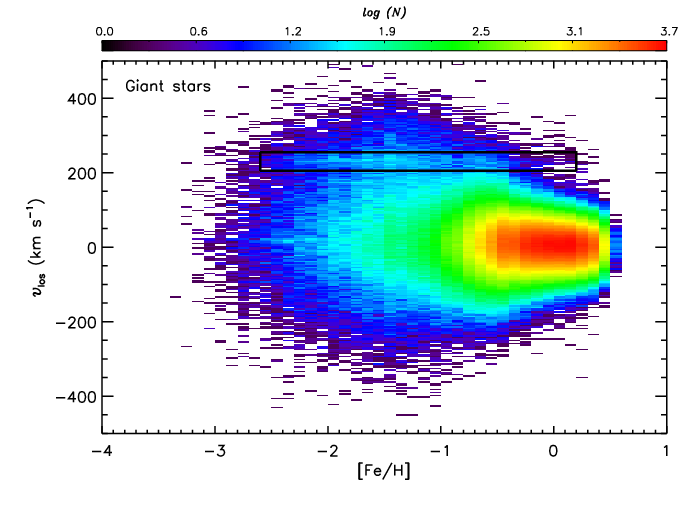}
\caption{Number-density distributions (on a logarithmic scale) for dwarf (left panel) and giant (right panel) stars in the $v_{\rm los}$ versus [Fe/H] plane for stars in the Final Sample; the top color bars indicate the number densities for bin sizes of 
of 4.0 km\,s$^{-1}$ in $v_{\rm los}$ and 0.1 dex in [Fe/H].
In the right panel, the significant excess along $v_{\rm los} \sim 230$\,km\,s$^{-1}$, marked with a black box, is largely due to the contribution from member stars of Omega Centauri ($\omega$ Cen).}
\end{center}
\end{figure*}

\begin{figure*}
\begin{center}
\includegraphics[scale=0.375,angle=0]{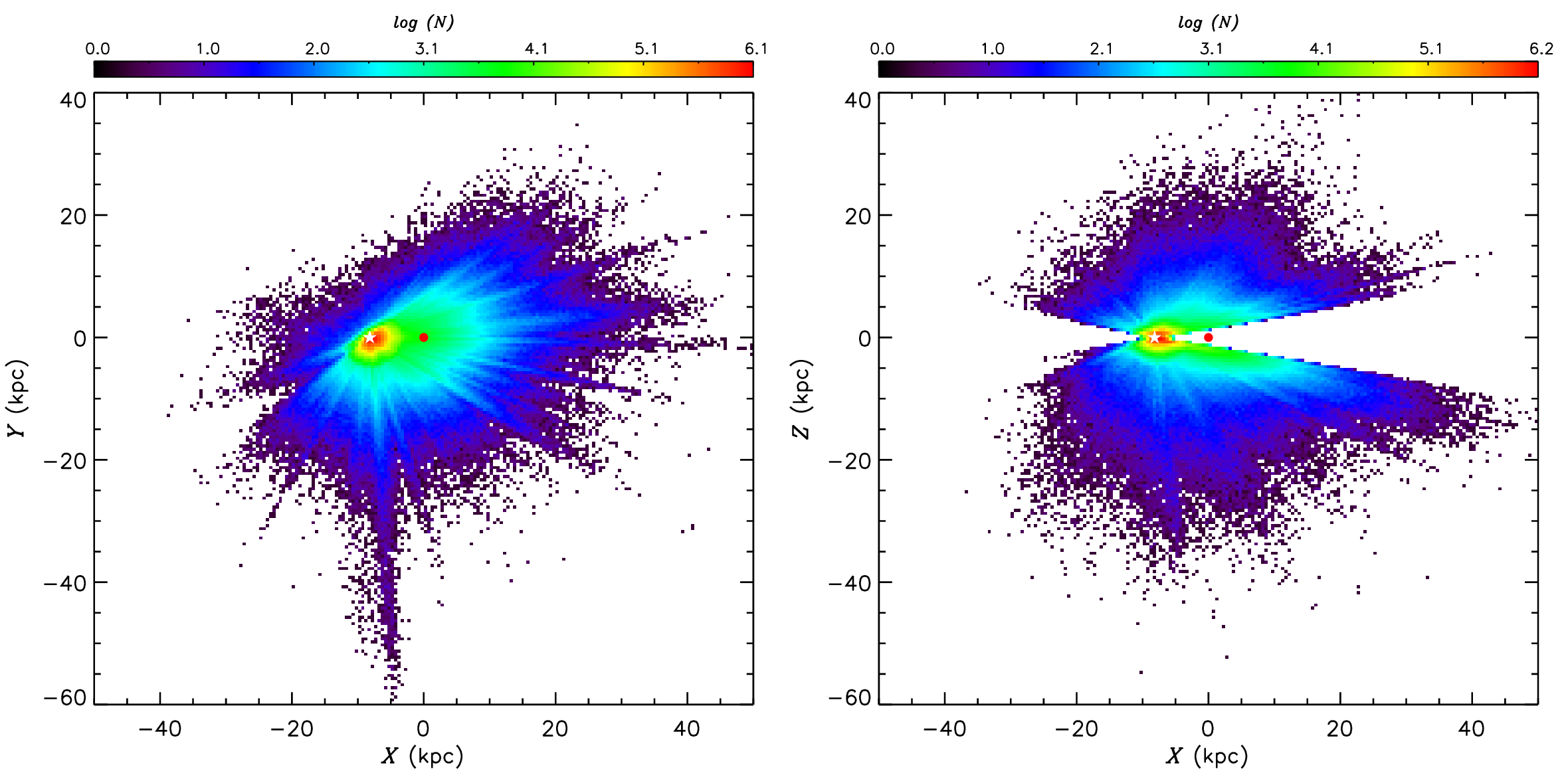}
\caption{Spatial number-density distributions for stars in the Final Sample in the $Y$ versus $X$ (left panel) and $Z$ versus $X$ (right panel) planes; the top color bars indicate the number densities for bin sizes of 0.5\,kpc in both axes.  The Sun is located at ($X$, $Y$, $Z$)\,=\,($-8.178$, 0.0, 0.0)\,kpc. The white star and the red dot mark the positions of the Sun and the Galactic center, respectively, in each panel.  The spike-like feature (especially visible in the left panel) is dominated by stars from the SMSS Main Survey fields (which go substantially deeper than the Shallow Survey fields), observed under good conditions and with low Galactic extinction.}
\end{center}
\end{figure*}

 \subsection{Comparison with APOGEE DR14 and DR16}

We now test the accuracy of our photometric-metallicity measurements by comparing our results with independent spectroscopic measurements from the APOGEE survey (Majewski et al. 2017).
Here we cross-match our Final Sample to APOGEE\,DR14 (Abolfathi et al. 2018) and DR16 (Ahumada et al. 2020), and found over 18,000 and 40,000 stars in common with spectral SNR greater than 50.
Comparisons between our photometric and APOGEE metallicities are shown in Fig.\,12.
Generally, our photometric metallicities agree very well with those of APOGEE DR\,14, with overall scatter around $0.13$\,dex, and negligible overall offsets of $0.00$\,dex and $-0.05$\,dex (photometric [Fe/H] minus APOGEE DR14 [Fe/H]) for dwarf and giant stars, respectively.
Moreover, no significant trends are found for the metallicity differences as a function of [Fe/H].
For APOGEE DR16, the results are similar, but with a moderate trend of the metallicity differences as a function of [Fe/H] for giant stars with [Fe/H]\,$< - 2.0$. 
We note that the [Fe/H] metallicity scale of APOGEE\,DR14 is calibrated by stellar clusters (Abolfathi et al. 2018; Holtzman et al. 2018), while no such external calibration is performed in APOGEE\,DR16 (Ahumada et al. 2020; J{\"o}nsson et al. 2020).
We speculate that the moderate trend found in the metallicity difference between our photometric estimates and the APOGEE\,DR16 spectroscopic estimates for giant stars possibly arises from this calibration issue.
We note the presence of a few outliers in Fig.\,12.  Their photometric-metallicity estimates may not be properly determined due to undetected binary and/or variable nature.
By assuming a typical uncertainty of 0.08\,dex for the metallicity measurement from APOGEE, the uncertainty of our photometric-metallicity estimate is about 0.05--0.15\,dex for both dwarf and giant stars with [Fe/H]\,$> -1.0$ and 0.10--0.20\,dex for giant stars  $-2.0 <$\,[Fe/H]\,$\le -1.0$, in excellent agreement with our internal test (see Fig.\,8).

As mentioned in Section\,2, the adopted photometry of SMSS DR2 is recalibrated by Huang et al. (2021).  As found in Huang et al. (2021), the photometric zero-points of $uv$-bands of the original SMSS DR2 exhibit significant trends with $E (B - V)$ (thus Galactic latitude) and declination.  
We thus examine the metallicity differences between photometric [Fe/H] and APOGEE\,DR16 [Fe/H] (due to the much larger sky coverage than DR14) with Galactic latitude and declination in Fig.\,13; the results show no significant trends.
Some weak/moderate variations in lower Galactic-latitude regions are possibly due to systematic errors of the extinction estimates in the SFD98 map (e.g., Schlafly et al. 2010; Yuan et al. 2013).
The relatively large scatters of the metallicity differences toward the low Galactic-latitude regions arise primarily from the large uncertainties of the extinction corrections in these regions in the SFD98 map.
This indicates that the photometric zero-point trends of the $uv$-bands in the original SMSS DR2  are corrected properly by Huang et al. (2021), and thus have only a minor effect on our estimated photometric metallicities. 

Finally, we examine our dwarf and giant star classifications, as defined in Fig.\,3, using APOGEE\,DR16. 
The surface gravity, log\,$g$, (from APOGEE DR16) distributions of the dwarf (black line) and giant (red line) stars, as classified by the current work, are shown in Fig.\,14.
Almost all the classified dwarf stars (19,712/19,953; 98.8 per cent) have log\,$g \ge 3.8$ and the giant stars (30,788/31,804; 96.8 per cent) have log\,$g < 3.8$, indicating the robustness of our dwarf and giant classifications. 

\subsection{Comparison with Metal-Poor Samples from the Literature}

In the above sections, the external checks focus on the metal-richer range of the 
photometric-metallicity estimates, i.e., [Fe/H]\,$> -1.0$ for dwarf stars and [Fe/H]\,$> -2.0$ for giant stars. In this section, we examine the more metal-deficient range of our photometric-metallicity estimates by comparing them to VMP ([Fe/H]\,$\le -2.0$) and EMP ([Fe/H]\,$\le -3.0$) samples from previous studies.

First, we compare the photometric-metallicity estimates for stars in our Final Sample to the small number of stars in common with the 1007 Pristine EMP star candidate samples (Starkenburg et al. 2017) with metallicity estimated from medium-resolution follow-up spectroscopy (Aguado et al. 2019). The typical uncertainty of the estimated spectroscopic metallicity is 0.2\,dex.
In total, 18 stars in common (10 dwarf and 8 giant stars) are found between our Final Sample and the EMP star candidate sample; the comparison is shown in the left panel of Fig.\,15.
Generally, in the VMP range, our photometric-metallicity estimates agree quite well with those from Aguado et al. (2019), with only a negligible offset of around $-0.02$\,dex (photometric minus spectroscopic), and a scatter of only 0.20\,dex.  Only one relatively large outlier in the EMP range is seen; it is known to be a carbon-enhanced metal-poor star ([C/Fe] = +1.08).

Secondly, our photometric metallicities are compared to those for low-metallicity candidates from the Best \& Brightest Survey (B\&B; Schlaufman \& Casey 2014), as reported by Limberg et al. (2021), which included estimates of [Fe/H] as well as other elemental-abundance ratios ([$\alpha$/Fe] and [C/Fe]) determined from low/medium-resolution spectra.
A total of $452$ stars (including 31 dwarf and 420 giant stars) with [Fe/H] $\le -1.0$ are found in common between the B\&B survey stars studied by Limberg et al. (2021) and our Final Sample; a comparison with these [Fe/H] estimates is presented in the middle panel of Fig.\,15.
Generally, our photometric estimates of [Fe/H] agree quite well with those from Limberg et al. (2021). Stars with reported enhancements in carbon ([C/Fe]\,$> +0.6$) are indicated with blue circles, and from inspection, they often have photometric-metallicity estimates that are higher than the spectroscopic estimates, as expected due to contamination of our metallicity-sensitive $uv$ bands from molecular carbon features. For the VMP range, the scatter is only 0.21\,dex (the scatter is similar if stars with [C/Fe]\,$> +0.6$ are discarded), and the offset is nearly zero. In the EMP range, about half of the stars are carbon-enhanced, and lie well above the one-to-one line, as expected.

Finally, we compare our photometric metallicities to spectroscopic estimates from the $R$-Process Alliance sample (RPA; Hansen et al. 2018; Sakari et al. 2018; Ezzeddine et al. 2020; Holmbeck et al. 2020) and the EMP star candidate sample (Jacobson et al. 2015; Marino et al. 2019) selected from the SkyMapper commissioning survey.  

The RPA sample contains over 600 metal-poor star candidates with metallicities estimated from follow-up high-resolution spectra.
Over 300 stars (including only five dwarf stars) are found in common between our Final Sample and the RPA sample; the comparison of [Fe/H] estimates is shown in the right panel of Fig.\,15.
Overall, the photometric-metallicity scale is in excellent agreement with that of the spectroscopic one in the metal-poor range, i.e., $-3.5 <$\,[Fe/H]\,$\le -1.0$, with a negligible offset of $-0.09$\,dex (photometric minus spectroscopic), and a scatter of 0.25\,dex (reduced to 0.24\,dex if stars with [C/Fe]\,$> +0.6$ are discarded). 
We note that the scatter of the metallicity difference for the VMP range is only 0.27\,dex, also with a negligible offset of $-0.06$\,dex.

Among the 139 EMP candidates from Jacobsen et al. (2015) and Marino et al. (2019), 99 stars (all giants) are included in our Final Sample; the comparison of [Fe/H] estimates is shown in the right panel of Fig.\,15.  We note that the spectroscopic metallicities from these authors were estimated by considering the effects of NLTE. 
Again, the photometric metallicities are consistent with the spectroscopic estimates, with a moderate offset of $-0.14$\,dex (photometric minus spectroscopic), and a scatter of 0.34\,dex (reduced to 0.25\,dex if stars with [C/Fe]\,$> +0.6$ are discarded).
The moderate offset found here is partially due to departures from the adopted LTE metallicity in our training set (which did not consider NLTE effects; see ealier discussion in Section\,3.2).
For the VMP range, the scatter is 0.33\,dex (reduced to 0.21\,dex if stars with [C/Fe]\,$> +0.6$ are discarded), with an offset of $-0.09$\,dex.

According to the above tests, the precision of our photometric-metallicity estimates is about 0.20--0.25\,dex for giant stars with [Fe/H]\,$\le -2.0$, by assuming a typical uncertainty of 0.15--0.20\,dex for the spectroscopic determinations. This result is consistent with our internal test in Section\,3.3.
The systematic offsets are typically within 0.10--0.15\,dex.

Finally, we note the recent work by Chiti et al. (2021), who presented photometric-metallicity estimates for 280,000 giant stars also from the SMSS\,DR2.
As shown in Fig.\,B1, their  metallicity estimates are quite consistent with our results, within expected errors.
 
 \begin{table*}
\centering
\caption{Sample Content}
\begin{tabular}{ccccccc}
\hline
\hline
&Dwarf &Giant&All\\
\hline
\textbf{Total}&\textbf{19,001,501}&\textbf{5,238,351}&\textbf{24,239,852}\\
\hline
\textbf{Stars with [Fe/H] measurements}&\textbf{19,001,501}&\textbf{5,238,351}&\textbf{24,239,852}\\
\text{[Fe/H]} measured by $u - G_{\rm BP}$&1,532,891&260,612&1,793,503\\
\text{[Fe/H]} measured by $v - G_{\rm BP}$&4,061,828&4,977,739&9,039,567\\
\text{[Fe/H]} measured by two colors&13,406,782&--&13,406,782\\
\hline
\textbf{Stars with $T_{\rm eff}$ measurements}&\textbf{19,001,501}&\textbf{5,238,351}&\textbf{24,239,852}\\
\hline
\textbf{Stars with distance measurements}&\textbf{15,356,014}&\textbf{5,178,665}&\textbf{20,534,679}\\
Distance estimated by Gaia EDR3 parallax&15,065,139&3,063,852&18,128,991\\
Distance estimated by color-absolute magnitude fiducials&290,875&2,114,813&2,405,688\\
\hline
\textbf{Stars with age measurements}&\textbf{14,984,221}&\textbf{3,027,520}&\textbf{18,011,741}\\
\hline
\textbf{Stars with RV measurements}&\textbf{1,219,405}&\textbf{1,101,128}&\textbf{2,320,533}\\
RV measured from Gaia-ESO DR3&5321&1338&6659\\
RV measured from GALAH DR3+&252,362&163,830&416,192\\
RV measured from APOGEE DR16&16,028&27,933&43,961\\
RV measured from Gaia DR2&66,095&865,606&1,525,701\\
RV measured from RAVE DR5&16,533&7494&24,027\\
RV measured from LAMOST DR7&201,399&31,027&232,366\\
RV measured from SDSS/SEGUE DR16&28,095&3642&31,737\\
RV measured from AEGIS&39,080&--&39,080\\
RV measured from B\&B&132&86&218\\
RV measured from  bibliographic collections&466&126&592\\
\hline
\end{tabular}
\end{table*}

\begin{table*}
\centering
\caption{A Naive Relation between [Fe/H] and [$\alpha$/Fe]}
\begin{tabular}{cccccccccc}
\hline
\hline
[Fe/H]&$-3.50$&$-2.00$&$-1.50$&$-1.00$&$-0.50$&$-0.25$&$0.00$&$+0.25$&$+0.50$\\
\hline
[$\alpha$/Fe]&$+0.80$&$+0.60$&$+0.40$&$+0.30$&$+0.20$&$+0.10$&$0.00$&$-0.20$&$-0.20$\\
 \hline
\end{tabular}
\end{table*}

 \section{Effective temperature}
In this section, we derive effective temperature ($T_{\rm eff}$) estimates for our sample stars.
To accomplish this, we first train metallicity-dependent $T_{\rm eff}$--color relations using the stars in common between LAMOST DR7 and SMSS DR2.
As shown by Huang et al. (2015b), the effective temperature scale derived from LAMOST by LASP is in excellent agreement with that of interferometric measurements for both dwarf and giant stars.
The selection of the training set is the same as described in Section\,3.1.
A total of 143,187 dwarf (log\,$g \ge 3.8$) and 28,655 giant (log\,$g < 3.8$) stars are selected for the training set.

To obtain the metallicity-dependent  $T_{\rm eff}$--color relations, a second-order 2-D polynomial (with 6 free parameters, including cross terms) is adopted to fit the data points for dwarf and giant stars in the training set separately:
 \begin{equation}
\theta_{\rm eff}  =  a_{0,0} + a_{0,1}y + a_{0,2}y^2 + a_{1,0}x +  a_{1,1}xy + a_{2,0}x^2\text{,}
 \end{equation}
 where $\theta_{\rm eff} = 5040/T_{\rm eff}$, and $x$ and $y$ represent  $(G_{\rm BP} - G_{\rm RP})_0$ and [Fe/H], respectively.
Three-sigma clipping is applied in the fitting process. The resultant fit coefficients are listed in Table 2.
The scatter of the fitting residual is 64\,K and 48\,K for dwarf and giant stars (see Fig.\,16), respectively.  From inspection of this figure, it is clear that the metallicity sensitivity is essentially gone below [Fe/H] = $-2.0$.

Using the derived metallicity-dependent $T_{\rm eff}$--color relations, values of $T_{\rm eff}$ are obtained for all sample stars, based on their stellar color  $(G_{\rm BP} - G_{\rm RP})_0$ and photometric [Fe/H].

 \section{Distance Estimates}
 \subsection{Distances Derived from Parallaxes}

By properly correcting for the parallax zero-points found by the official {\it Gaia} team (Lindegren et al. 2021), and adopting a sophisticated set of direction-dependent priors, Bailer-Jones et al. (2021) obtain distance estimates from the {\it Gaia} EDR3 parallax measurements for 1.47 billion stars.
Tests from both mock data and open clusters suggest that the derived distances are reliable out to several kpc.
For stars with reliable parallax measurements (i.e., relative parallax error smaller than 25\%, parallax greater than 0.167\,mas, and renormalized unit weight error smaller than 1.4) from {\it Gaia} EDR3, we thus adopt the (geometric) distances estimated by Bailer-Jones et al. (2021) directly. In total, 18,128,991 stars (see Table\,3) have their distances estimated in this manner.

With the geometric distances from Bailer-Jones et al. (2021), we derive the $G$-band absolute magnitudes of stars by correcting for extinction from the map of SFD98 (corrected for a 14\% systematic over-estimate in the map).
The resulting H-R diagram is shown in Fig.\,17 (here over 12 million stars with relative parallax error smaller than 10 per cent and parallax greater than 0.4\,mas are shown).
With the help of PARSEC isochrones (Bressan et al. 2012; Marigo et al. 2017), we further empirically divide the dwarf stars  into turn-off, main-sequence, and binary stars.
The turn-off stars are defined by significant absolute magnitude variations with stellar age (guided by the PARSEC isochrones in Fig.\,17).
The binary stars are those with absolute magnitudes above the most metal-rich isochrones with [M/H]\,$= +0.5$.
The remaining objects in the dwarf star region are defined as main-sequence stars. 

 \subsection{Distances Derived from Metallicity-Dependent Color-Absolute Magnitude Fiducials}
 \subsubsection{Dwarf Stars} 
 To derive the distances for dwarf stars without distance estimates from Bailer-Jones et al. (2021), we define a metallicity-dependent color-absolute magnitude relation.
 For this, we fit the absolute magnitude $M_{G_0}$, as a function of color $(G_{\rm BP} - G_{\rm RP})_0$ and photometric [Fe/H], using over 1.5 million main-sequence stars defined in Fig.\,17 with two additional cuts: Galactic latitude $|b| \ge 20^{\circ}$ (to reduce the uncertainty from reddening corrections), and metallicity uncertainty smaller than 0.30\,dex.
To obtain the relation, a third-order 2-D polynomial (with 10 free parameters, including cross terms) is adopted to fit $M_{G_0}$ as a function of $(G_{\rm BP} - G_{\rm RP})_0$ and [Fe/H]:
 \begin{equation}
 \begin{split}
M_{G_0} & =  a_{0,0} + a_{0,1}y + a_{0,2}y^2 + a_{0,3}y^3 + a_{1,0}x + \\
 & a_{1,1}xy + a_{1,2}xy^2 + a_{2,0}x^2 + a_{2,1}x^2y + a_{3,0}x^3\text{,}
 \end{split}
 \end{equation}
where $x$ and $y$ represent $(G_{\rm BP} - G_{\rm RP})_0$ and [Fe/H], respectively.
Three-sigma clipping is applied in the fitting process.
The resultant fit coefficients are listed in Table\,2.
The fitting residuals of the absolute magnitude as a function of $(G_{\rm BP} - G_{\rm RP})_0$ and [Fe/H] are shown in Fig.\,18, in the left and right panels, respectively.
Typically, the scatter of the residuals is about 0.15 to 0.20\,mag.

Using this relation, we now can derive absolute magnitude (and thus distance estimates) for those main-sequence stars without parallax-based distance determinations from their color $(G_{\rm BP} - G_{\rm RP})_0$ and [Fe/H].
To exclude turn-off stars (absolute magnitude depends on age), a color cut $(G_{\rm BP} - G_{\rm RP})_0 \ge 1.0$ is adopted.
In this manner, distances are derived for a total of 290,875 main-sequence stars.
Finally, we note that a small fraction of turn-off/binary stars are incorrectly assigned  distances in this way, but their distances are unavoidably underestimated.

 \subsubsection{Giant Stars}
For giant stars without distance estimates from {\it Gaia} EDR3, we adopted the same method used in Huang et al. (2019), to which we refer the interested reader,  to derive absolute magnitudes (thus distance estimates) using their stellar color $(g - i)_0$, photometric [Fe/H], and the empirical color-magnitude fiducials interpolated from six globular clusters.
As shown by Huang et al. (2019), the typical uncertainty of the derived distance by this method is about 16\%,  without significant systematic errors.
In total, distances are obtained by this method for 2,114,813 giant stars.

To summarize, 85\% (i.e., over 20 million) stars of our Final Sample have distance estimates derived either from the {\it Gaia} EDR3 parallax or from the metallicity-dependent color-absolute magnitude fiducials.

 \section{Age Estimates}
In this section, we derive age estimates for stars in our Final Sample with accurate absolute magnitude estimates from {\it Gaia} EDR3 (see Section\,5.1).
For this, we adopt the conventional Bayesian approach, similar to that of J{\o}rgensen \& Lindegren (2005), Xiang et al. (2017), and Huang et al. (2020), by matching the observed stellar parameters with theoretical isochrones.
The evolution of a star is largely determined by age $\tau$, initial stellar mass $M$, and chemical composition $Z$.
According to the Bayesian approach, the (posterior) probability density function of the three parameters can be written as:
  \begin{equation}
f (\tau, M, Z) = Nf_0 (\tau, M, Z)L (\tau, M, Z)\text{,}
 \end{equation}
 where $f_0$ is  the  prior  probability  distribution function  of  the  three parameters, $L$ is the  likelihood  distribution function, and $N$ is a normalization  factor  to  make $\iiint f_0 (\tau, M, Z)d\tau dMdZ = 1$.
The likelihood distribution function $L$ is given by:
 \begin{equation}
 L (\tau, M, Z) = (\prod_{i = 1}^{n}\frac{1}{\sqrt{2\pi}\sigma_{i}}) \times \exp{(\frac{-\chi^2}{2})}\text{,}
 \end{equation}
where the $\chi^2$ is:
  \begin{equation}
\chi^2 = \sum_{i = 1}^{n} (\frac{O_i - P_i (\tau, M, Z)}{\sigma_i})^2\text{.}
 \end{equation}
 Here $O$ denotes the observed stellar parameters $(G_{\rm BP} - G_{\rm RP})_0$, $M_{G_0}$, [M/H], and $P$ denote the theoretical values from isochrones at a given $\tau$, $M$, and $Z$.
The number of observed parameters is $n$, and $\sigma_i$ is the error of the $i$th observed parameter.
 
For the  prior probability  distribution function, we assume a flat metallicity and age distribution since they are not well known.
For the stellar mass, we adopt a Salpeter initial mass function (Salpeter 1955), thus:
\begin{equation}
f_0 (\tau, M, Z) \propto m^{-2.35}\text{.}
\end{equation}
 
For the theoretical model, the PARSEC isochrones (Bressan et al. 2012; Marigo et al. 2017) are chosen for age estimates.
For this purpose, we produce a grid of 0.1-15.2\,Gyr isochrones in steps of 0.2 Gyr for age\,$< 1.2$\,Gyr and of 0.5\,Gyr for age\,$>1.2$\,Gyr, and 0.02\,dex in [M/H] ($-2.2 <$\,[M/H]\,$< +0.5$).
The full grids include over $1.3 \times 10^6$ individual models.
Finally, the photometric [Fe/H] is converted to [M/H] for comparison using the formula:
\begin{equation}
\text{[M/H]} = \text{[Fe/H]} + \text{log} (0.694\times10^{\text{[}\alpha\text{/Fe]}} + 0.306)\text{,}
\end{equation}
from Salaris \& Cassisi (2005). 
For the conversion, we adopt a naive relation between [Fe/H] and [$\alpha$/Fe] (e.g., Venn et al. 2004), as given in Table\,4.
For individual stars, their values of [$\alpha$/Fe] are then simply interpolated from the above relation and their values of [M/H] are further obtained by Equation\,10.

For each star, we then obtain a posterior probability distribution function (PDF) of stellar ages given by our Bayesian approach.
The final age of a star is given by the median of the resulted posterior PDF, and its uncertainty is defined as the half of the difference between the 84 and 16 per cent values of the resulting posterior PDF.

In total, we estimate stellar ages for over 18 million stars with accurate absolute magnitude estimates from {\it Gaia} EDR3.
However, as shown by the isochrones in Fig.\,17, this method is mainly effective in estimating ages for turn-off stars; the uncertainties of the derived ages in other regions of the H-R diagram are very large.
For ease of use of this sample, we provide type classifications as defined in Fig.\,17 for those stars with accurate $M_{G_0}$.
Finally, we compare the stellar ages derived here to those from Sanders \& Das (2018; hereafter SD18).
In SD18, isochrone ages are derived for 3 million stars by using the observed constraints from {\it Gaia} DR2 (Gaia Collaboration et al. 2018) and spectroscopic surveys (i.e., APOGEE, Gaia-ESO, GALAH, LAMOST, RAVE, and SDSS/SEGUE).
Over 120,000 main-sequence turn-off stars (with type classified `TO' in Fig.\,17 and $3.6 \le$\,log\,$g \le 4.1$ from SD18) with relative age uncertainty smaller than 30\% and relative parallax errors smaller than 15\% are found in common with SD18; a comparison of the stellar age determinations between this work and  SD18 is shown in Fig.\,19.
Generally, the derived age in the current work agrees very well with that of SD18, with essentially zero offset of the relative age difference (${\text{age}_{\rm TW} - \text{age}_{\rm SD18}}$)/${\text{age}_{\rm SD18}}$ and a scatter of the relative age difference of around 19\%.

\begin{table*}
\centering
\caption{Description of the Final Sample}
\begin{tabular}{lll}
\hline
\hline
Field&Description&Unit\\
\hline
SMSS\_ID&Unique ID for the SMSS catalog&--\\
sourceid&Cross-matched Gaia EDR3 source ID&--\\
ra&Right Ascension from SMSS DR2 (J2000)&degrees\\
dec&Declination from  SMSS DR2 (J2000)&degrees\\
gl&Galactic longitude derived from ICRS coordinates&degrees\\
gb&Galactic latitude derived from ICRS coordinates&degrees\\
u/v/g/r/i/z&PSF magnitudes for the six SkyMapper bands from SMSS DR2&--\\
uc/vc/gc/rc&Corrected PSF magnitudes for the SkyMapper uvgr-bands by Huang et al. (2021)&--\\
err\_u/v/g/r/i/z&Uncertainties of PSF magnitudes for the six SkyMapper bands from SMSS DR2&mag\\
G\_C/BP/RP&Magnitudes for the thee {\it Gaia} three bands EDR3; note G\_C represents a calibration-corrected G magnitude&--\\
err\_G\_C/BP/RP&Uncertainties of magnitudes for the three {it Gaia} bands from EDR3&mag\\
ebv\_sfd&Value of $E (B - V)$ from from the extinction map of SFD98, corrected for a 14\% systematic&--\\
BR$_0$/uB$_0$/vB$_0$&Intrinsic colors of $(G_{\rm BP} - G_{\rm RP})_0$, $(u - G_{\rm BP})_0$, and $(v - G_{\rm BP})_0$&-- \\
err\_BR$_0$/uB$_0$/vB$_0$&Uncertainties of intrinsic colors of $(G_{\rm BP} - G_{\rm RP})_0$, $(u - G_{\rm BP})_0$, and $(v - G_{\rm BP})_0$&mag\\
\text{[Fe/H]}&Photometric metallicity&--\\
\text{err\_[Fe/H]}&Uncertainty of photometric metallicity&dex\\
\text{[Fe/H]\_flg}&Flag to indicate the stellar color(s) used in estimating [Fe/H], which takes the values ``ub", ``vb", and ``ub+vb"&--\\
$T_{\rm eff}$&Effective temperature&K\\
err\_$T_{\rm eff}$&Uncertainty of effective temperature&K\\
\text{dist\_adop}&Distance&kpc\\
\text{err\_dist\_adop}&Uncertainty of distance&kpc\\
\text{dist\_adop\_flg}&Flag to indicate the method used to derive distance, which takes the values ``parallax", ``CAF", and ``NO"&--\\
X/Y/Z&3D positions in the right-handed Cartesian system&kpc\\
err\_X/Y/Z&Uncertainties of 3D positions in the right-handed Cartesian system&kpc\\
$R_{\rm GC}$&Galactocentric distance&kpc\\
err\_$R_{\rm GC}$&Uncertainty of Galactocentric distance&kpc\\
$R$&Projected Galactocentric distance onto the Galactic plane&kpc\\
err\_$R$&Uncertainty of projected Galactocentric distance&kpc\\
\text{age}&Stellar age&Gyr\\
\text{err\_age}&Uncertainty of stellar age&Gyr\\
rv\_adop&Radial velocity&km s$^{-1}$\\
err\_rv\_adop&Uncertainty of radial velocity&km s$^{-1}$\\
rv\_adop\_flg&Flag to indicate the source of radial velocity, which takes the values ``Gaia-ESO", ``GALAH", ``APOGEE'', ``Gaia",&--\\
& ``RAVE", ``LAMOST", ``SEGUE", ``AEGIS", ``B\&B", ``LIT" and ``NO"&--\\
parallax&Parallax from {\it Gaia} EDR3&mas\\
err\_parallax&Uncertainty of parallax from {\it Gaia} EDR3&mas\\
pmra&Proper motion in Right Ascension direction from  {\it Gaia} EDR3&mas yr$^{-1}$\\
err\_pmra&Uncertainty of proper motion in Right Ascension direction from  {\it Gaia} EDR3&mas yr$^{-1}$\\
pmdec&Proper motion in Declination direction  from  {\it Gaia} EDR3&mas yr$^{-1}$\\
err\_pmdec&Uncertainty of proper motion in Declination direction from  {\it Gaia} EDR3&mas yr$^{-1}$\\
$V_R$/$V_{\phi}$/$V_Z$&3D velocity in the Galactocentric  cylindrical system&km s$^{-1}$\\
err\_$V_R$/$V_{\phi}$/$V_Z$&Uncertainty of 3D velocity in the Galactocentric  cylindrical system&km s$^{-1}$\\
ruwe&Renormalised unit weight error from  {\it Gaia} EDR3&--\\
type&Flag to indicate classifications of stars, which takes the values ``dwarf" and ``giant" &--\\
subtype&Flag to indicate further sub-classifications of dwarf stars, which takes the values  ``TO", ``MS", ``Binary" and ``NO"&--\\
\hline
\end{tabular}
\end{table*}

 \section{Radial velocities}
For stars in the Final Sample, we collected their radial-velocity measurements from previous spectroscopic surveys, including Gaia-ESO DR3 (Gilmore et al. 2012; Randich \& Gilmore 2013), GALAH DR3+ (Buder et al. 2021), SDSS DR16 (Ahumada et al. 2020), Gaia DR2 (Katz et al. 2019), RAVE DR5 (Kunder et al. 2017), LAMOST DR7\footnote{\url{http://dr7.lamost.org/}}, AEGIS (Yoon et al. 2018), B\&B survey (Schlaufman \& Casey 2014), as reported by Limberg et al. (2021), and several bibliographic collections (Barbier-Brossat\& Figon 2000; Malaroda et al. 2006; de Bruijne et al. 2012).
The radial-velocity measurement from the higher-resolution survey is adopted if a star has measurements from two or more different surveys.
The zero points of the radial velocities yielded by different surveys (except the B\&B survey and the bibliographic collections) are all calibrated to those given by the APOGEE radial-velocity standard stars (Huang et al. 2018). Details of the adopted radial velocities are provided in Table\,3.
In total, over 2.3 million stars in our sample have radial-velocity measurements from those surveys.
The  stellar number distributions of $v_{\rm los}$ versus [Fe/H] for dwarf (left panel) and giant (right panel) stars are shown in Fig.\,20.
From inspection, these plots exhibit clear trends of decreasing radial-velocity dispersion with increasing [Fe/H], as expected from the Galaxy's formation and evolution history.
In the right panel of Fig.\,20, the member stars (marked with a black box) with constant radial velocity of $\sim 230$\,km\,s$^{-1}$ (e.g., Harris et al. 2010) and a large metallicity range (at least from $-2.5$ to $-0.5$; e.g., Villanova et al. 2004) of Omega Centauri ($\omega$ Cen) are clearly seen.

 \section{The Final Sample and Perspectives}
 We summarize the Final Sample content in this section.
 This sample contains over 19 million dwarf and 5 million giant stars with metallicity estimated from the stellar colors of SMSS DR2 and {\it Gaia} EDR3 (see Section\,3), and effective temperature estimates from  $(G_{\rm BP} - G_{\rm RP})_0$ and photometric [Fe/H] (see Section\,4).
 Among the over 24 million sample stars, over 20 million of them have distance estimates (see Section\,5), and over 18 million of them have age estimates (see Section\,6).
We also include astrometric information from {\it Gaia} EDR3 (Gaia Collaboration et al. 2021). 

From those stars with distance measurements, 3D positions in the right-handed Cartesian system ($X$, $Y$, $Z$, with $X$ toward the direction opposite to the Sun, $Y$ in the direction of Galactic rotation, and $Z$ toward the north Galactic pole) and Galactocentric cylindrical system ($R$, $\phi$, $Z$; with $R$ the projected Galactocentric distance, increasing radially outwards, $\phi$ in the direction of Galactic rotation, and $Z$ the same as that in the Cartesian system) are calculated.
Spatial distributions of our sample stars in the $Y$--$X$ and $Z$--$X$ planes are shown in Fig.\,21.
Finally, for stars with distances, proper motions and  radial velocities, we derive 3D velocities in Cartesian, Galactocentric cylindrical, and Galactocentric spherical systems.
The three velocity components are represented by ($U$, $V$, $W$) in the Cartesian system, ($v_R$, $v_{\phi}$, $v_Z$) in the Galactocentric cylindrical system, and ($v_r$, $v_{\theta}$, $v_{\phi}$) in the Galactocentric spherical system. 
In the Galactocentric spherical system, $r$ is the Galactocentric distance, increasing radially outwards, $\theta$ increases toward the south Galactic pole, and $\phi$ is in the direction of Galactic counter-rotation.
In above calculations, we  set  the  Solar Galactocentric distance to $R_0 = 8.178$\,kpc (Gravity Collaboration et al. 2019), its total azimuthal velocity to $v_{\phi, \odot} = 247$\,km\,s$^{-1}$ (Reid et al. 2009), and adopt the Solar motion with respect to the Local Standard of Rest ($U_{\odot}$, $V_{\odot}$, $W_{\odot}$)\,$=$\,(7.01, 10.13, 4.95)\,km\,s$^{-1}$ (Huang et al. 2015a).

Table\,5 lists the columns contained in the Final Sample catalog.  
Compared to other previous work based on SMSS DR1.1 or DR2 (e.g., Casagrande et al. 2019; Huang et al. 2020; Chiti et al. 2021), this work presents a sample including the largest numbers of stars and derived stellar properties to date. 
As we have demonstrated with multiple comparisons to external spectroscopy for stars in common, the range and accuracy of the estimated photometric metallicity for the stars in our sample are the best yet achieved, essentially limited by the precision of the input photometry (which will improve once data from the SMSS Main Survey are available).  

Finally, we remark that this sample will be very powerful for various Galactic studies, including (but not limited to):

 \begin{enumerate}[label=\arabic*)]

\item Characterization of the properties of stellar populations, and the kinematics and structure of the Galactic thin/thick discs and the stellar halo

\item Identification of tidal streams and debris from disrupted dwarf galaxies and globular clusters, and constraining their formation and evolution history

\item Studies of the perturbation and heating history of the Galactic disk system

\item Probing the gravitational potential and dark matter distribution of the Galaxy

\end{enumerate}

All of the above will ultimately advance our knowledge of the assembly history of our Galaxy, and other large spiral galaxies.
The Final Sample catalog will be publicly available within three months after publication (will be released at National Astronomical Data Center of China: \url{https://doi.org/10.12149/101073}), and can be shared earlier upon reasonable requests to the lead author.

 \section{Summary}
In this comprehensive work, we present precise photometric-metallicity estimates for some 24 million stars (over 19 million dwarf and 5 million giant stars) based on stellar colors from SMSS DR2 and {\it Gaia} EDR3, using training data sets with spectroscopic metallicity measurements from either high-resolution (PASTEL and SAGA) or low/medium-resolution spectroscopic surveys (LAMOST and SDSS/SEGUE).
From consideration of a variety of external tests, the typical uncertainty of our photometric-metallicity estimates is about 0.05--0.15\,dex for both dwarf and giant stars with [Fe/H]\,$> -1.0$, 0.10--0.20\,dex for giant stars with $-2.0 <$\,[Fe/H]\,$\le -1.0$, and 0.20--0.25\,dex for giant stars with [Fe/H]\,$\le -2.0$.
Our techniques appear useful down to metallicities [Fe/H] $\sim -3.5$.
Effective temperature ($T_{\rm eff}$) estimates for all of the sample stars are determined by metallicity-dependent $T_{\rm eff}$--color relations, constructed from a training set of 1.6 million stars in common between LAMOST and SMSS.
Distances for over 85\% of the sample stars (i.e., over 20 million) are obtained either from  the {\it Gaia} EDR3 parallax-based measurements by  Bailer-Jones et al. (2021) or from our own estimates based on metallicity-dependent color-absolute magnitude fiducials.
For stars with accurate absolute magnitude determined from {\it Gaia} EDR3 parallaxes, stellar ages are further estimated for over 18 million sample stars by matching their observed properties to those from theoretical isochrones.

Astrometric information from {\it Gaia} EDR3 is also included in this sample.
Finally, over 9.5\% of our sample stars have radial-velocity measurements from previous  spectroscopic surveys.

The number of stars with metallicity estimates in the current work  is 4--5 times larger than the number obtained by the largest spectroscopic survey conducted to date (LAMOST).
This huge dataset will provide vital constraints on understanding the formation and evolution of our Galaxy, and serves to demonstrate the power of ongoing and future photometric stellar surveys.\\
\\
 \section*{Acknowledgements} 
We would like to thank the referee for helpful comments.
It is a pleasure to thank Dr. Deokkeun An for a thorough read of the manuscript and helpful comments.
This work is supported by National Key R\&D Program of China No. 2019YFA0405500 and National Natural Science Foundation of China grants 11903027, 11833006, 11973001, 11603002, 11811530289 and U1731108. 
Y. H. is supported by the Yunnan University grant C176220100006. 
We used data from the European Space Agency mission Gaia (\url{http://www.cosmos.esa.int/gaia}), processed by the Gaia Data Processing and Analysis Consortium (DPAC; see \url{http://www.cosmos.esa.int/web/gaia/dpac/consortium}). 
T.C.B. acknowledges partial support from grant PHY 14-30152, Physics
Frontier Center/JINA Center for the Evolution of the
Elements (JINA-CEE), awarded by the US National Science
Foundation. His participation in this work was initiated by conversations that took place during a visit to China in 2019, supported by a PIFI Distinguished Scientist award from the Chinese Academy of Science.
Y.S.L. acknowledges support from the National Research Foundation (NRF) of Korea grant funded by the Ministry of Science and ICT (NRF-2018R1A2B6003961
and NRF-2021R1A2C1008679).
CAO acknowledges support from the Australian Research Council through Discovery Project DP190100252.

The national facility capability for SkyMapper has been funded through ARC LIEF grant LE130100104 from the Australian Research Council, awarded to the University of Sydney, the Australian National University, Swinburne University of Technology, the University of Queensland, the University of Western Australia, the University of Melbourne, Curtin University of Technology, Monash University and the Australian Astronomical Observatory. SkyMapper is owned and operated by The Australian National University's Research School of Astronomy and Astrophysics. The survey data were processed and provided by the SkyMapper Team at ANU. The SkyMapper node of the All-Sky Virtual Observatory (ASVO) is hosted at the National Computational Infrastructure (NCI). Development and support the SkyMapper node of the ASVO has been funded in part by Astronomy Australia Limited (AAL) and the Australian Government through the Commonwealth's Education Investment Fund (EIF) and National Collaborative Research Infrastructure Strategy (NCRIS), particularly the National eResearch Collaboration Tools and Resources (NeCTAR) and the Australian National Data Service Projects (ANDS).

The Guoshoujing Telescope (the Large Sky Area Multi-Object Fiber Spectroscopic Telescope, LAMOST) is a National
Major Scientific Project built by the Chinese Academy of Sciences. Funding for the project has been provided by the
National Development and Reform Commission. LAMOST is operated and managed by the National Astronomical Observatories, Chinese Academy of Sciences.

\appendix
\section{Extinction Coefficients}
In this section, we investigate the variations of extinction coefficients of $(G_{\rm BP} - G_{\rm RP})_0$ for $R (u - G_{\rm BP})$,  $R (v- G_{\rm BP})$,  $R (G_{\rm BP} - G_{\rm RP})$, and $R_G$ with intrinsic stellar color $(G_{\rm BP} - G_{\rm RP})_0$, using model spectra from Husser et al. (2013).
To do this, the extinction coefficients of $(G_{\rm BP} - G_{\rm RP})_0$ for $R (u - G_{\rm BP})$,  $R (v- G_{\rm BP})$,  $R (G_{\rm BP} - G_{\rm RP})$, and $R_G$ are predicted by an $R_V = 3.1$ Fitzpatrick extinction law at $E (B - V ) = 0.7$ for source spectra of different effective temperature, $T_{\rm eff}$ (with log\,$g = 4.5$ and Solar metallicity).
In total, 33 groups of extinction coefficients are calculated for sources with $T_{\rm eff}$ from 3600\,K to 6800\,K in steps of 100\,K.
The resulting extinction coefficients, as a function of intrinsic stellar color $(G_{\rm BP} - G_{\rm RP})_0$, are shown in Fig.\,A1.
From inspection, significant variations are found for the extinction coefficients with $(G_{\rm BP} - G_{\rm RP})_0$.
To quantitively describe those variations, third-order polynomial fits were performed, and the obtained relations are shown in the top part of each panel of Fig.\,A1.

\begin{figure}
\begin{center}
\includegraphics[scale=0.33,angle=0]{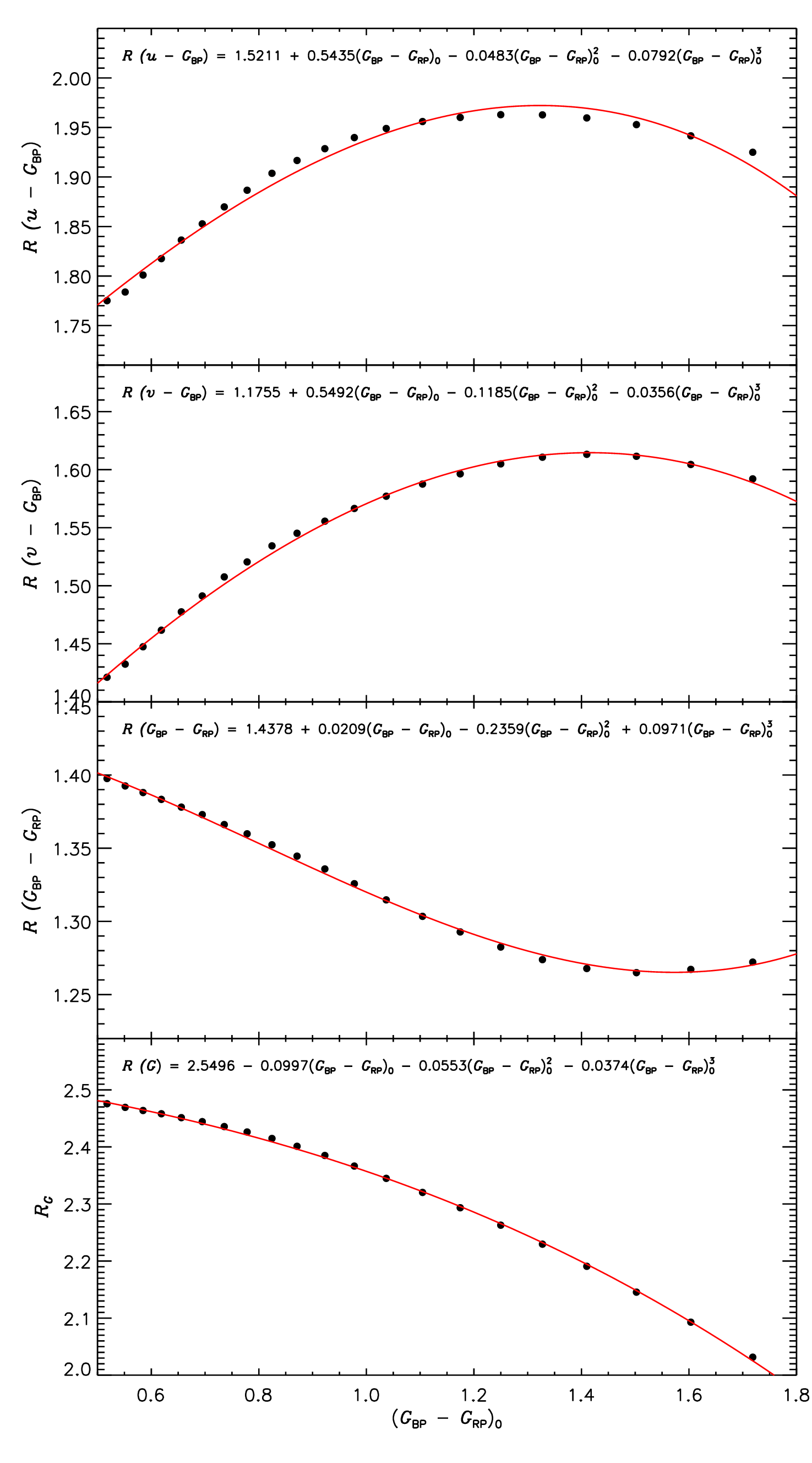}
\caption{Extinction coefficients, as a function of intrinsic stellar color $(G_{\rm BP} - G_{\rm RP})_0$, for $R (u - G_{\rm BP})$,  $R (v- G_{\rm BP})$,  $R (G_{\rm BP} - G_{\rm RP})$, and $R_G$, from top to bottom, respectively. Red solid lines show the fits given in each panel.}
\end{center}
\end{figure}

\section{Comparison with Chiti et al. (2021)}
Most recently, Chiti et al. (2021) have presented a sample of photometric-metallicity estimates for about 280,000 giant stars with [Fe/H] $\ge -0.75$ from SMSS DR2.
Almost all of their stars also have metallicity estimates in our sample; we show the metallicity comparison in Fig.\,B1.
Generally, their metallicity estimates are in good agreement with our results, with a scatter of 0.28\,dex and a median offset of only 0.02\,dex.

\begin{figure}
\begin{center}
\includegraphics[scale=0.33,angle=0]{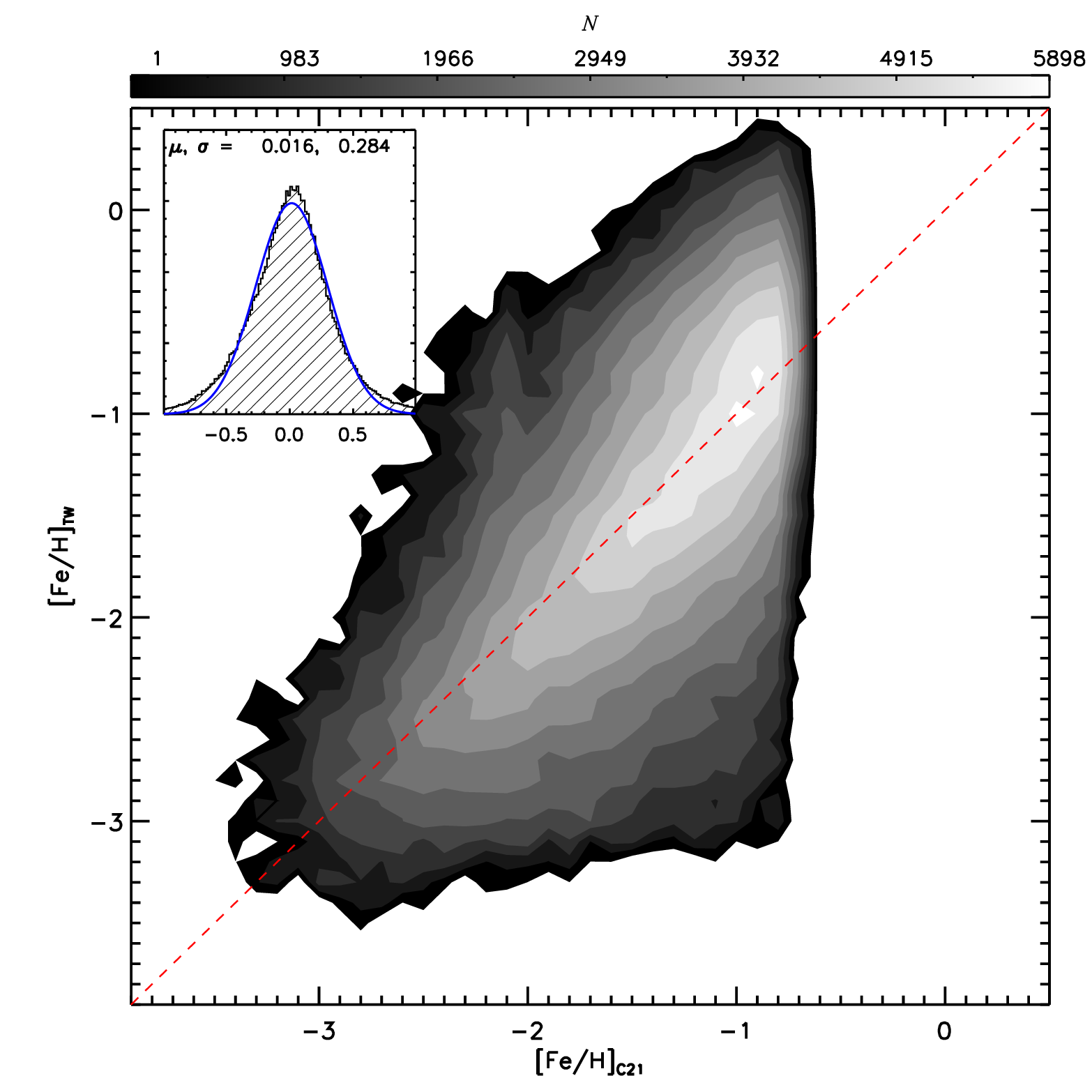}
\caption{Comparison of estimated photometric metallicity, [Fe/H], from this work with that derived by Chiti et al. (2021; C21) for about 270,000 stars in common. The color-coded contour of the stellar number density is shown.  
In the top-left corner, the distribution of differences in derived [Fe/H] between the C21 values and our estimates are shown. 
The blue line is a Gaussian fit to the distribution of the residuals, with the mean and dispersion of the Gaussian marked in the plot.}
\end{center}
\end{figure}

\end{document}